# Pragmatical physics-based model for ladle lifetime prediction

By Stein Tore Johansen[1*], Bjørn Tore Løvfall[1] and Tamara Rodriguez Duran[2]

[1]    SINTEF; Stein Tore Johansen, Bjørn Tore Løvfall

[2]    Sidenor; Tamara Rodriguez Duran

*    Correspondence: Stein Tore Johansen; stein.t.johansen@sintef.no

*Abstract*

In this paper we develop a physics-based model for the erosion of lining in steel ladles. The model predicts the temperature evolution in the liquid slag, steel, refractory bricks and outer steel casing. The flows of slag and steel is due to forced convection by inert gas injection, vacuum treatment (extreme bubble expansion), natural convection and waves caused by the gas stirring. The lining erosion is due to dissolution of refractory elements into the steel or slag. The mass and heat transfer coefficients inside the ladle, during gas stirring, is modelled based on wall functions which take the distribution of wall shear velocities as a critical input. The wall shear velocities are obtained from CFD (Computational Fluid Dynamics) simulations for sample of scenarios, spanning out the operational space, and using curve fitting a model could be built. The model is capable of reproducing both thermal evolutions and erosion evolution well. Deviations between model predictions and industrial data are discussed.  The model is fast and has been tested successfully in a "semi-online" application. The model source code is made available to the public on https://github.com/SINTEF/refractorywear.

## Introduction

In the steel industry ladles are frequently used to keep, process or transport steel. Ladles are designed to typically hold metal masses ranging from 80 to 300 tons (Figure 1). The melt typically consists of high temperature liquid steel and some slag, which when interacting inner wall of the ladle will harm the wall integrity and lead to significant wear. In order to reduce the wear, temperature resistant and chemically resistant refractory brick are applied to build an inner barrier, typically three layers of wear bricks (inner lining) which should last for a long time in contact with the liquid steel, and at the same time protect the ladle from showing hot areas.. In this paper we will address the inner lining erosion at a Sidenor plant. In this case the ladle application is Secondary Metallurgy (SM).

 During SM many processes may going on. The SM ladles have installed a porous plug at the bottom. Gas (Ar of N) injected through the plug is responsible for liquid steel stirring. The rising flow of the liquid steel promotes the inclusion decantation from the steel to the slag, and homogenizes the temperature and chemical composition.

The main objective of the SM is to obtain the correct chemical composition and have enough temperature for the casting process. In addition, there are several important tasks which must be complete during the SM, as for example inclusion and gases removal. In order to reach these objectives, Sidenor has a SM mill consisting of two Ladle furnaces (LF) and a Vacuum Degasser (VD). Each of the LFs have three electrodes, which are responsible of heating the slag, steel and ferro-additions. The ladle contains the steel and the slag for all the production process from the EAF to the end of the casting process. The liquid steel has a temperature of around 1700 K in the ladle, and it is covered with slag. The slag avoids the contact between the steel and the atmosphere, has lower density than steel and consists basically of lime and oxide elements. The slag conditioning can be improved during the SM by adding slag-formers.

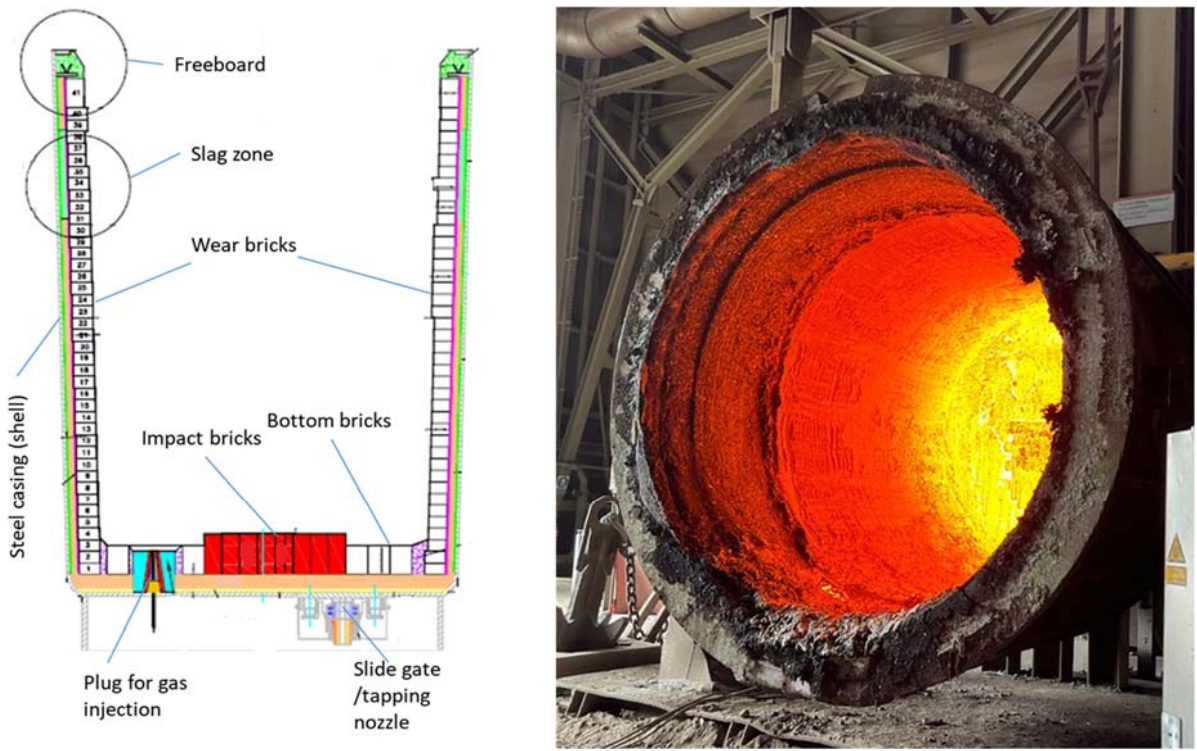

*Figure 1 Left: Sketch of cross section of a typical steel ladle, with wear refractory bricks, permanent lining (between wear bricks and steel casing), steel casing, bottom bricks, bottom plug for bottom gas blowing and slide gate for transfer into casting tundish. Right: Hot ladle that has been in use and is waiting for the next heat. Maximum steel capacity is around 150 tons.*

In order to handle the liquid steel and slag with such high temperature, the ladle is built with a strong outer steel shell whose inside is covered with layers of insulating materials (refractory). The refractory is made of ceramic and its most important properties are:
  i.   handle the high temperature
  ii.  favorable thermal properties
  iii. high resistance against erosion when in contact with liquid steel and slag

The inner layer of refractory bricks, which are in contact with the liquid steel, are eroded by the interaction with the hot metal and the slag. Each heat erodes away the refractory bricks, and after several heats, they are so eroded that it is not safe to use the ladle one more time. The refractory is visually checked after each heat and depending of its state, the ladle may be used one more heat, put aside for repair or demolished. In case of repair, the upper bricks of the ladle, which are more eroded due to slag chemical impact, will be replaced by new ones. Once the ladle is repaired, it is taken back into production. Later, based on continuing visual inspection, the ladle may be deemed ready for demolition. In this case, the entire inner lining is removed and relined with new bricks.

One important goal for Sidenor is to reduce the refractory costs by finding new methods for extending the refractory life. One of the key points is to use the same ladle during more heats without compromising the safety, but another important issue is to understand better the mechanism that drives the refractory erosion, in order to avoid as much as possible the worst working practices and so to enlarge the working life.

### *Target for the development*

The main goal is to develop a model whose results can help to decide whether the ladle could work one more heat safely. The model should exploit both historic and current production data. The model added to the knowledge of the operators could be exploited and contribute to cognitive elements of the model.

In addition, the model should give information about which parameters dominates the ladle refractory erosion and give tips about which precautions may be taken to extend the refractory lifetime.

### **Previous works on ladle lining erosion**

In the past many works have been published, dealing with properties of refractory bricks ((Mahato et al., 2014),(Wang et al., 2015)), advising on improvement to produce high quality bricks. A more general review of MgOC refractories was given by Kundu and Sarkar (Kundu and Sarkar, 2021). The corrosion-erosion mechanisms have been studied in a few papers ((Kasimagwa et al., 2014), (Jansson, 2008), (Mattila et al., 2002), (Huang et al., 2013), (LMMGROUP, 2020), (Zhu et al., 2018)). In the opinion of these authors, the most thorough approach was given by Zhu et al. (Zhu et al., 2018). Bai et al. (Bai et al., 2022) investigated the impact of slag penetration into the MgOC bricks.

In order to predict erosion of the refractory both temperatures, fluid compositions and mass transfer mechanisms bust be in place. The heat balance was studied is some specialized works ((Çamdali and Tunç, 2006),(Glaser et al., 2011), (Zimmer et al., 2008), (Duan et al., 2018), and (Zhang et al., 2009)). The effects of slag composition was studies in multiple works ( (Bai et al., 2022; Jansson, 2008; Kasimagwa et al., 2014; Mattila et al., 2002; Sarkar et al., 2020; Sheshukov et al., 2016; Zhu et al., 2018)). A critical step in developing prediction models is the local mass transfer lining and slag metal. This mass transfer has this far been treated by semiempirical models ( [11], [17], (Wang et al., 2022)). In one work 3D computational fluid dynamics was applied (Wang et al., 2022), and where predictions seems to agree with observations. However, they did not report the diffusivities used in their model and the underlying erosion-corrosion models were empirical and tuned to the data. It was found that these tuning factors would depend on the operating conditions.

In the industry the refractory wear is known to be a result of i) Thermal stresses, ii) Dissolution of the refractory bricks into slag/metal, and iii) dissolution of the binder materials into slag/metal. In addition, mechanical stresses imposed on the refractory during cleaning operations will impact on erosion and lifetime. In addition to these multiple phenomena, several others are involved (WanHaoRefractory, 2023).

The impact of thermal stresses will be most severe at the bottom of the ladle when hot steel meets colder refractory. As the velocity of the metal at the moment of impact is high, this is where we expect the maximum thermal stresses. The colder the ladle wall is when it meets hot steel at high speed, the larger is the risk of cracks formation on the ladle wall.

It must be noted that time between heats have significant effects on thermo-stress induced erosion. The temperature distribution on the ladle refractory wall at the filling time is an important parameter that can be predicted using the model to be presented below. However, the addition of a heating burner at the ladle waiting station is not included for now.

## The pragmatism-based approach to a model for ladle lining erosion

We have previously defined a methodology "Pragmatism in Industrial Modelling" (J. Zoric et al., 2015; Johansen and Ringdalen, 2018; Johansen, Stein Tore et al., 2017; Zoric, Josip et al., 2015) which is especially suited for developing fast and sufficiently accurate industrial models. In a twin paper (Johansen et al., 2023) we have outlined the methodology that was applied in this work and the learning that may be exploited in future projects. Here we explain the details of the physics-based model.

The objective of the model is to be able to advice or support operators in assessing if it is safe to use the refractory in one more heat. In such an application, the erosion state of the refractory must be updated from heat to heat and simulation for a next virtual heat could be performed. The virtual heat should then contain as much information as possible about the next heat. The result of such a simulation and visual or optical inspection of the lining would then lay the foundation for the final assessment.

## Model simplifications and assumptions

The pragmatic model must be fast as we wish to simulate a transient ladle operation, lasting in the order of two hours, in less than a minute. This is critical as we wish to simulate all ladle operations within a year in a few hours in order to be applied directly in production, do tuning, or do parameter sensitivity analysis.

Figure 2 gives some ideas about the phenomena involved. The heating elements (electrodes) can be submerged in the slag, or work from above. They produce electric arcs that heat the liquid steel. The flow of the slag and liquid steel is not only a function of the gas flow rate applied for blowing, but also is influenced by several effects, such as the mass of steel and slag, vacuum pressure and the thermo-physical properties of the fluids.

The ladles are 3D objects, but due to speed requirements some overall model simplifications were done:

i) Model is 2D (cylinder symmetrical) with the porous bottom plug placed in the center. As a consequence, we assume that the gas/steel/slag flows can be seen as rotationally symmetric
ii) The stirring gas is inert (only provides mixing)
iii) In the side walls only the radial heat balance is included
iv) In the bottom only vertical heat balance is included
v) Solubility of MgO in the slag and solubility of C in the steel are assumed constant.
vi) The metal and the slag phases are stratified and are assumed to be internally perfectly mixed. The phases exchange mass and energy with each other and the refractory
vii) Above the slag energy is exchanged by radiation only
viii) Refractory erosion due to thermomechanical stresses is not considered

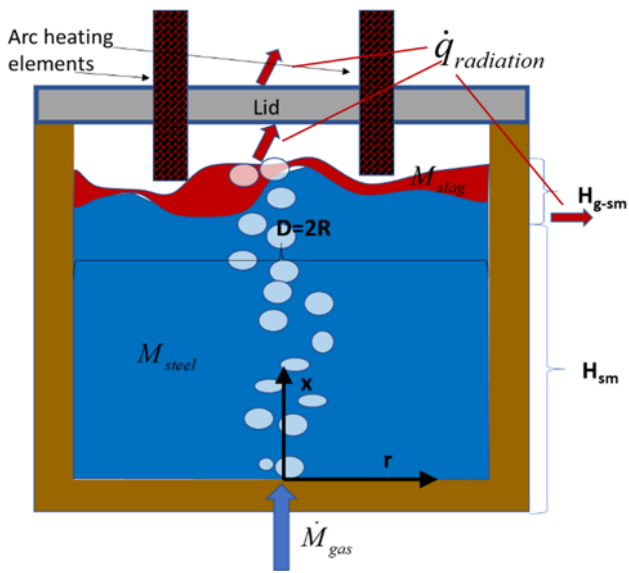

*Figure 2 Idealized, simplified ladle, with slag (red), metal(blue), gas bubbles, heating elements and refractory (brown)*

## Volumes and mass balances

As the model will compute situations with different amounts of steel and slag in the ladle, we have to take into account all these possible situations. Now the total volume of the slag and metal is represented by

$$V_{tot} = V_{steel} + V_{slag} = \alpha_{steel} V_{tot} + \alpha_{slag} V_{tot} \tag{1}$$

Accordingly, the mass of liquids inside the ladle is:

$$M_{tot} = \rho_{steel}\alpha_{steel}V_{tot} + \rho_{slag}\alpha_{slag}V_{tot} \qquad [2]$$

In our first approach, we neglect the volumes of the protruding impact element at the bottom and the volumes modified by eroded bricks. In this case, we have that the height of the metal-slag interface is positioned at height

$$H_{sm} = \alpha_{steel}V_{tot}/(\pi R^2), \qquad [3]$$

and the thickness of the slag layer is:

$$H_{g-sm} = \alpha_{slag}V_{tot}/(\pi R^2) \qquad [4]$$

The mass balance for the ladle must also be respected. That is for the slag

$$\frac{dM_{slag}}{dt} = \dot{M}_{slag,EAF} - \dot{M}_{slag,tapped} + \sum_{k=1}^{N_{slag}} \dot{m}_{slag,k} \qquad [5]$$

Here $\dot{M}_{slag,EAF}$ and $\dot{M}_{steel,EAF}$ are the transient mass flow rates of slag and steel coming into the ladle during tapping from the EAF. $\dot{m}_{slag,k}$ is the mass flow rate of added slag former of type k. Typically a slag former of type k, total mass $m_{slag,k}$, can be assumed to be added during one numerical time step, between time $t^n$ and $t^{n+1}$, such that

$$M_{slag}^{n+1} = M_{slag}^n + \Delta t(\dot{M}_{slag,EAF} - \dot{M}_{slag,tapped}) + m_{slag,k} \qquad [6]$$

For the metal we have:

$$\frac{dM_{steel}}{dt} = \dot{M}_{steel,EAF} - \dot{M}_{steel,tapped} + \sum_{k=1}^{N_{alloy}} \dot{m}_{alloy,k} \qquad [7]$$

$\dot{M}_{slag,tapped}$ and $\dot{M}_{steel,tapped}$ are the transient mass flow rates of slag and steel tapped out of the ladle. Similarly, $\dot{m}_{alloy,k}$ is the mass flow rate of added alloy of type k. As for the slag, an alloy of type k, total mass $m_{alloy,k}$, can be assumed to be added during one numerical time step, between time $t^n$ and $t^{n+1}$, such that

$$M_{steel}^{n+1} = M_{steel}^n + \Delta t(\dot{M}_{steel,EAF} - \dot{M}_{steel,tapped}) + m_{alloy,k} \qquad [8]$$

Based on the equations [5]-[8], the phase densities, the purge gas fractions present in each phase, and corrections for the eroded ladle radius, we can compute the transient interface position for the metal and slag interfaces.

## Thermal model
### Ladle walls

The ladle side wall is built with a number of radial layers, as shown in Figure 3. Next, we let the numerical grid, as seen the figure, represent each vertical layer of wear bricks, and stack multiple layers on top of each other to represent the entire side wall of the ladle. The colors in Figure 3 represent different properties of the materials. The bottom part of the refractory is built of a stack of disks, which also may be represented by Figure 3, but now rotated 90 degree clockwise.

In this manner, the numerical grid for the ladle wall and casing temperature will consist of one one-dimensional grid (here 7 cells) for the bottom and N one-dimensional grids for the vertical wall (Nx7 cells). For the horizontal and radial heat balance we have

$$\frac{\partial}{\partial t}\left(\rho C_p T^w\right)_i = \frac{\partial}{r \partial r}(\lambda r \frac{\partial T^w}{\partial r}) \qquad [9]$$

Equation [9] is discretized for each layer according to

$$\underbrace{2\pi \Delta x_i r_k \Delta y_k}_{\Delta V} \left(\rho C_p\right) \frac{T_{i,k}^{w,n+1} - T_{i,k}^{w,n}}{\Delta t} = \\ 2\pi \Delta x_i \left(\lambda_k^+ r_k^+ \left(T_{i,k+1}^{w,n+1} - T_{i,k}^{w,n+1}\right) - \lambda_k^- r_k^- \left(T_{i,k}^{w,n+1} - T_{i,k-1}^{w,n+1}\right)\right) \qquad [10]$$

Above + and – represent the value at the positive and negative sides of the cell-face. $\Delta x_i$ is the vertical height of the grid cell at level i cell while $r_k$ is the radial position index for the cell. We use harmonic averages for the cell-face thermal conductivities $\lambda_{k+1}^-$ and $\lambda_k^+$

$$\lambda_k^+ = \lambda_{k+1}^- = \frac{2\lambda_{k+1}}{\Delta y_{k+1}} \frac{2\lambda_k}{\Delta y_k} / (\frac{2\lambda_{k+1}}{\Delta y_{k+1}} + \frac{2\lambda_k}{\Delta y_k}) \qquad [11]$$

and where $r_k$ is defined according to Figure 3:

$$r_k^+ \equiv r_{k+1}^- = r_k + \Delta y_k / 2 \equiv r_{k+1} - \Delta y_{k+1} / 2 \qquad [12]$$

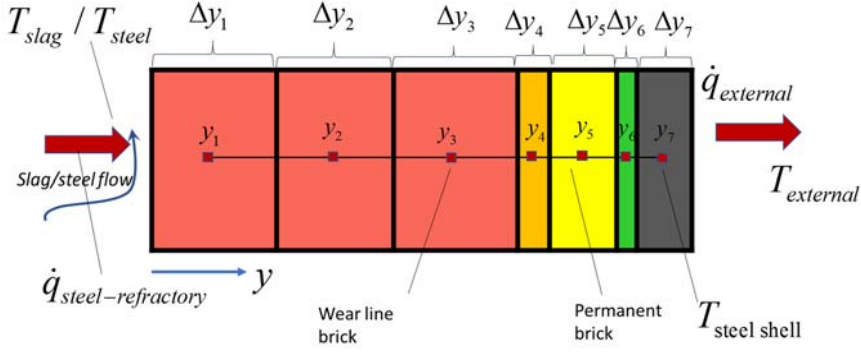

*Figure 3 Element of the refractory where the transient thermal heat balance is addressed.*

In the cell contacting the hot liquid steel and slag (k=1) we have

$$\underbrace{2\pi \Delta x_i r_1 \Delta y_1}_{\Delta V}(\rho C_p)\frac{T_{i,1}^{w,n+1}-T_{i,1}^{w,n}}{\Delta t} = 2\pi \Delta x_i \begin{pmatrix} \lambda_j^+ r_j^+ \left(T_{i,2}^{w,n+1}-T_{i,1}^{w,n+1}\right)- \\ r_1^- \begin{pmatrix} \alpha_{i,metal}\tilde{h}_i^{metal,inner}\left(T_{metal}-T_{i,1}^{w}\right)+ \\ \alpha_{i,slag}\tilde{h}_i^{slag,inner}\left(T_{slag}-T_{i,1}^{w}\right)+ \\ \alpha_{i,gas}\tilde{h}_i^{radiation}\left(T_{lid}-T_{i,1}^{w}\right) \end{pmatrix} \end{pmatrix} \quad [13]$$

In eq. [13] $\alpha_{i,metal}$, $\alpha_{i,slag}$ and $\alpha_{i,gas}$ are the local volume fractions of the phases contacting the element $\Delta x_i$ at a given time.

$$\tilde{h}_i^{metal,inner} = \frac{2\lambda_{i,1}}{\Delta y_1}\tilde{h}_i^{metal,flow} / (\frac{2\lambda_{i,1}}{\Delta y_1}+\tilde{h}_i^{metal,flow}) \quad [14]$$

$$\tilde{h}_i^{slag,inner} = \frac{2\lambda_{i,1}}{\Delta y_1}\tilde{h}_i^{slag,flow} / (\frac{2\lambda_{i,1}}{\Delta y_1}+\tilde{h}_i^{slag,flow}) \quad [15]$$

$$\tilde{h}_i^{gas,inner} = \frac{2\lambda_{i,1}}{\Delta y_1}\tilde{h}_i^{rad} / (\frac{2\lambda_{i,1}}{\Delta y_1}+\tilde{h}_i^{rad}) \approx \frac{2\lambda_{i,1}}{\Delta y_1} \quad [16]$$

Where the external temperature is given by $T_{EXT}$. The radiation heat transfer coefficient is given by

$$\tilde{h}_i^{rad} = \sigma\varepsilon_i(T_{i,1,w}^{2}+T_{EXT}^{2})(T_{i,1,w}+T_{EXT}) \quad [17]$$

and where the wall temperature is further approximated the temperature in the near wall cell at the previous time step:

$$\tilde{h}_i^{rad} = \sigma\varepsilon_i(T_{i,1}^{w,n2}+T_{EXT}^{2})(T_{i,1}^{w,n}+T_{EXT}) \quad [18]$$

For the outer wall at $y_{NJ} = y_7$ (steel casing) we have:

$$\underbrace{2\pi\Delta x_i r_{NJ} \Delta y_{NJ}}_{\Delta V}(\rho C_p)\frac{T_{i,NJ}^{w,n+1} - T_{i,NJ}^{w,n}}{\Delta t} = 2\pi\Delta x_i \left(r_{NJ}^+ \left(\tilde{h}_i^{ext}\left(T_{EXT} - T_{i,1}^w\right)\right) - \lambda_{NJ}^- r_{NJ}^- \left(T_{i,NJ}^{w,n+1} - T_{i,NJ-1}^{w,n+1}\right)\right) \quad [19]$$

Here the external heat transfer coefficient is estimated by a sum of natural convection and radiation. The convective external heat transfer coefficient $h_{NC}$ is given by equation [104] using the properties for air. The dimension used in the convective model should be the half height of the ladle standing straight up. The effective external heat transfer coefficient is then

$$\tilde{h}_i^{ext} = \sigma\varepsilon_{casing}(T_{i,1}^{w2} + T_{EXT}^2)^2(T_{i,1}^w + T_{EXT}) + h_{NC} \quad [20]$$

When the ladle is located inside a cabinet, with the ladle kept inside a compartment with external walls, the effective heat emissivity in equation [20] can be multiplied by a factor of 0.5.

It should be noted that the external heat transfer coefficients must be adjusted to the situation the ladle experiences (melt refining, transport to casting station, casting, transport to waiting station, waiting). If the external heat transfer conditions varies between the different events this must be handled in an appropriate manner such that we can tune the model to get a realistic thermal history for the ladle.

The model for the bottom energy is completely analogous to what is described above, but now with the discrete equation

$$\underbrace{\pi R^2 \Delta x_m}_{\Delta V}(\rho C_p)\frac{T_m^{b,n+1} - T_m^{b,n}}{\Delta t} =$$
$$\pi R^2 \left(\lambda_m^+ \left(T_{m+1}^{b,n+1} - T_m^{b,n+1}\right) - \lambda_m^- \left(T_m^{b,n+1} - T_{m-1}^{b,n+1}\right)\right) \quad [21]$$

Here R is the inner radius of the ladle. For the element close to the liquid steel (we assume that steel flows into the bottom at time = 0.0 sec) we have:

$$\underbrace{\pi R^2 \Delta x_{m=NM}}_{\Delta V}(\rho C_p)\frac{T_{m=NM}^{b,n+1} - T_{m=NM}^{b,n}}{\Delta t} =$$
$$\pi R^2 \left(\tilde{h}_{steelflow-bottom}\left(T_{steel} - T_{NM}^{b,n+1}\right) - \lambda_1^+ \left(T_{NM}^{b,n+1} - T_{NM-1}^{b,n+1}\right)\right) \quad [22]$$

$$h_{steelflow-bottom} = \frac{2\lambda_{NM,j}}{\Delta x_{NM}}\tilde{h}_i^{metal,flow} / (\frac{2\lambda_{NM,j}}{\Delta x_{NM}} + \tilde{h}_i^{metal,flow}) \quad [23]$$

*For the bottom element (steel shell) we have:*

$$\underbrace{\pi R^2 \Delta x_{m=1}}_{\Delta V}\left(\rho C_p\right)\frac{T^{b,n+1}_{m=1}-T^{b,n}_{m=1}}{\Delta t}=\pi R^2\left(\lambda_1^+\left(T_2^{b,n+1}-T_1^{b,n+1}\right)-\tilde{h}_{bottom}\left(T_1^{b,n+1}-T_{EXT}\right)\right) \qquad [24]$$

where we estimate

$$\tilde{h}_{bottom} \approx 10.0 \text{ W/(m}^2\text{K)} \qquad [25]$$

### Radiation - Wall temperatures and heat transfer above the slag/metal

Above the liquid phase the refractory will only see the top lid, the other parts of the wall and the metal surface. We will assume that the top lid is adiabatic, such that no energy is drained out through the lid. We now have to assess radiation transfer between different inner wear bricks and the top surface of the slag/metal. The radiation flux out from a surface with emissivity $\varepsilon_p$ and temperature $T_p$ is given by

$$q_{rad}=\varepsilon_p \sigma T_p^4 \qquad [26]$$

The radiation heat flow from surface elements $A_1$ to $A_2$ is given by ("View factor," 2022)

$$\widetilde{Q_{1\rightarrow 2}}=\varepsilon_1 \sigma T_1^4 \int_{A_1}\int_{A_2}\frac{\cos\theta_1 \cos\theta_2}{\pi s^2}dA_2 dA_1 \qquad [27]$$

The geometrical configuration is seen from Figure 4.

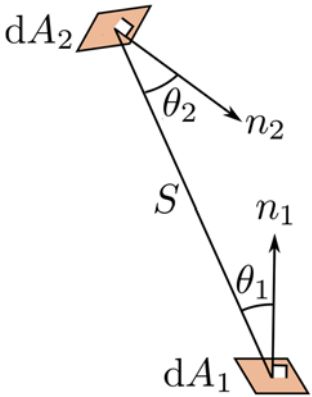

*Figure 4 Geometrical arrangement for radiation exchange between areas $A_1$ and $A_2$ ("View factor," 2022).*

The radiation heat flow from $A_2$ to $A_1$ is then

$$\widetilde{Q_{2\rightarrow 1}}=\varepsilon_2 \sigma T_2^4 \int_{A_1}\int_{A_2}\frac{\cos\theta_1 \cos\theta_2}{\pi s^2}dA_2 dA_1 \qquad [28]$$

The heat flow between the two surfaces A₁ and A₂ can be given by (Goodman, 1957):

$$\dot{Q}_{2\to 1} = -\dot{Q}_{1\to 2} = \frac{\left(T_2^4 - T_1^4\right)}{1/\varepsilon_2 + 1/\varepsilon_1 - 1}\sigma \int_{A_1}\int_{A_2} \frac{\cos\theta_1 \cos\theta_2}{\pi s^2} dA_2 dA_1 \qquad [29]$$

Based on equations [26]-[29], the surface normal vectors $\mathbf{n}_1$ and $\mathbf{n}_2$ and the vector connecting area elements $dA_1$ and $dA_2$, all radiation heat flows can be computed. These are $\dot{Q}_{w,m\to slag-metal}$ (from brick number m to slag-metal interface), $\dot{Q}_{w,m\to ceiling}$ (from brick number m to ceiling), and $\dot{Q}_{slag-metal\to ceiling}$ (from slag-metal interface to ceiling). The direct radiation between bricks was ignored. As the radiation from the slag-metal interface must respect that the slag only covers a fraction $\alpha_{slag}$ of the total free surface area. Hence, the radiation temperature $T_{slag-metal}^4$ is replaced by:

$$T_{slag-metal}^4 = \alpha_{slag} T_{slag}^4 + (1-\alpha_{slag}) T_{metal}^4 \qquad [30]$$

It is further assumed that ceiling (ladle lid) is adiabatic and that the slag and metal is well mixed. However, for the refractory bricks the thermal conduction heat flux into the inner wall surface brick and the net radiation flux must balance. The surface temperature of the wall bricks are then given as

$$T_{wall-surface,k} \approx \frac{\dfrac{2\lambda_k}{\Delta y_k} T_{wall,k}^n + T_{slag-metal}^n \Psi \left(T_{wall,k}^n + T_{slag-metal}^{2,n}\right)\left(T_{wall,k}^n + T_{slag-metal}^n\right)}{\left\{\Psi\left(T_{wall,k}^2 + T_{slag-metal}^{2,n}\right)\left(T_{wall,k}^n + T_{slag-metal}^n\right) + \dfrac{2\lambda_k}{\Delta y_k}\right\}} \qquad [31]$$

This illustrated the fact that it is surface temperature that communicates radiation and not volume averaged temperature for the computational cell.

The factor $\Psi$ is given by

$$\Psi_k = \frac{\sigma}{1/\varepsilon_{brick} + 1/\varepsilon_{slag-metal} - 1} \frac{\left(R(1-1/\sqrt{2})\right)\max\left(0; x_k - x_{slag-metal}\right)\cdot \dfrac{R^2}{2}\Delta\theta}{\pi\left\{\left(R(1-1/\sqrt{2})\right)^2 + \left(\max\left(0; x_k - x_{slag-metal}\right)\right)^2\right\}^2}, \qquad [32]$$

where $R\Delta\theta\Delta x$ is the vertical area element ($\dfrac{R^2}{2}\Delta\theta$) on which the computation is made.

The heat flows may be converted to heat transfer coefficient by rewriting equation [29] as

$$\dot{Q}_{2\to 1} = \left[\frac{\sigma}{1/\varepsilon_2 + 1/\varepsilon_1 - 1}\int\int_{A_1 A_2}\frac{\cos\theta_1 \cos\theta_2}{\pi s^2}dA_2\right]dA_1\left(T_2^4 - T_1^4\right) =$$

$$dA_1\left\{\frac{\sigma}{1/\varepsilon_2 + 1/\varepsilon_1 - 1}\int\int_{A_1 A_2}\frac{\cos\theta_1 \cos\theta_2}{\pi s^2}dA_2\left(T_2^2 + T_1^2\right)\left(T_2 + T_1\right)\right\}\left(T_2 - T_1\right) \quad , \qquad [33]$$

$$\equiv dA_1 \tilde{h}_{2\to 1}\left(T_2 - T_1\right)$$

where $\tilde{h}_{2\to 1}$ is the heat transfer coefficient expressed by the $\{\ \}$ above.

As the lid is adiabatic we have the following condition to fulfil:

$$Q_{slag-metal\to ceiling} + \sum_{w,m} Q_{w,m\to ceiling} = 0 \quad , \qquad [34]$$

From [34] we compute the ceiling temperature.

Effective heat transfer coefficient

The effective heat transfer coefficient $\tilde{h}_{liq}$, in the liquid steel and slag, may now be estimated based on three different contributions: 1. The wave induced contribution $\tilde{h}_{wave}$, elaborated in *Appendix E Wave induced heat transfer*, 2. The heat transfer due to bubble stirring $\tilde{h}_{stirring}$, is elaborated in *Appendix F Inner wall heat transfer coefficients due to forced convection bydue to bubble stirring*, and 3. The heat transfer due to natural convection $h_{NC}$, is elaborated in Appendix D. Pure natural and effective convection heat transfer:

$$\tilde{h}_{liq} = \tilde{h}_{wave} + \left(\tilde{h}_{stirring}^{1/2} + \tilde{h}_{NC}^{1/2}\right)^2 \qquad [35]$$

*Heat balance for the slag*

Due to melting of additives (slag formers, refining substances, alloying elements) we have selected to represent the energy by the specific enthalpy h.

First, we give the slag enthalpy by a simplified relation:

$$H_{slag} = M_{slag}h_{slag}(T) = m_{slag,EAF}h_{slag,EAF} + \sum_{k}^{N_{slag}} m_{slag,k}h_{slag,k} =$$

$$m_{slag,EAF}C_{p,slag,EAF}T + \sum_{k=1}^{N_{slag}} m_{slag,k}\begin{cases} C_{p,s,k}^{slag}T; T \leq T_{k,1} \\ C_{p,s,k}^{slag}T_1 + \Delta h_k \dfrac{T - T_{k,1}}{T_{k,1} - T_{k,2}}; T_{k,1} < T < T_{k,2} \\ C_{p,s,k}^{slag}T_1 + \Delta h_k + C_{p,l,k}^{slag}\left(T - T_{k,2}\right); T \geq T_{k,2} \end{cases} \qquad [36]$$

Here the enthalpy for the solids are represented by $C_{p,s}T$, and for the liquids it is given by $C_{p,s}T_1 + \Delta h + C_{p,l}(T - T_2)$, where $C_{p,l}$ is the liquid heat capacity and $\Delta h$ is the heat of transforming the solid into a liquid state. The temperatures $T_1$ and $T_2$ are the temperatures where the phase transition (melting) starts and is completed, respectively.

The heat balance for the slag is then

$$\frac{d}{dt}(M_{slag}h_{slag}) = \sum_{i=1}^{NI} 2\pi R \Delta x_i \alpha_{i,slag} \tilde{h}_i^{slag,inner}(T_{i,1}^w - T_{slag}) + \dot{Q}_{slag}$$
$$+ \dot{M}_{slag,EAF} C_{p,slag} T_{slag,EAF} - \dot{M}_{slag,tapped} h_{slag}$$
$$+ \pi R^2 \tilde{h}_i^{slag,lid}(T_{lid} - T_{slag}) + \pi R^2 \tilde{h}_i^{slag,metal}(T_{steel} - T_{slag})$$
$$+ \sum_{k}^{N_{slag}} \dot{m}_{slag,k} h_{slag,k}(T_{feed})$$

[37]

Here $\alpha_{i,slag}$ is the slag fraction contacting brick number i and varies with time. $T_{feed}$ is the temperature of the materials at time of feeding, typically less than 100 °C. $\dot{M}_{slag,EAF}$ is the time dependent mass flow of slag coming from the EAF. $\tilde{h}_i^{slag,lid}$ is the heat transfer coefficient for slag surface – top lid heat exchange, and $\tilde{h}_i^{slag,metal}$ is the area averaged heat transfer coefficient between the metal and slag. $\dot{Q}_{slag}$ is the heating power supplied to the slag [W/kg]. All these quantities are in general varying with time.

By applying the mass balance [5] into [37] we obtain:

$$M_{slag}\frac{dh_{slag}}{dt} = \sum_{i=1}^{NI} 2\pi R \Delta x_i \alpha_{i,slag} \tilde{h}_i^{slag,inner}(T_{i,1}^w - T_{slag}) + \dot{Q}_{slag}$$
$$+ \dot{M}_{slag,EAF}(C_{p,slag} T_{slag,EAF} - h_{slag})$$
$$+ \pi R^2 \tilde{h}_i^{slag,lid}(T_{lid} - T_{slag}) + \pi R^2 \tilde{h}_i^{slag,metal}(T_{steel} - T_{slag})$$
$$+ \sum_{k}^{N_{slag}} \dot{m}_{slag,k} \{h_{slag,k}(T_{feed}) - h_{slag}\}$$

[38]

We may note that eq. [38] tells that the slag components fed at low temperature $T_{feed}$, will lower the enthalpy of the slag as $h_{slag,k}(T_{feed}) - h_{slag} < 0$.

*Heat balance for the metal*

As for the slag, the metal enthalpy $H_{steel}$ can be expressed by the specific enthalpy $h_{steel}$:

$$H_{steel} = M_{steel} h_{steel}(T) = m_{steel} h_{steel} + \sum_{k}^{N_{alloy+1}} m_{alloy,k} h_{alloy,k} =$$

$$m_{steel} C_{p,steel} T + \sum_{k=1}^{N_{alloy}} m_{alloy,k} \begin{cases} C_{p,s,k}^{steel} T ; T \leq T_{k,1} \\ C_{p,s,k}^{steel} T_{k,1} + \Delta h_k \dfrac{T - T_{k,1}}{T_{k,1} - T_{k,2}} ; T_{k,1} < T < T_{k,2} \\ C_{p,s,k}^{steel} T_{k,1} + \Delta h_k + C_{p,l,k}^{steel} \left( T - T_{k,2} \right) ; T \geq T_{k,2} \end{cases} \quad [39]$$

Similarly, for the metal (steel) we have

$$\begin{aligned} M_{steel} \frac{dh_{steel}}{dt} &= \sum_{i=1}^{NI} 2\pi R \Delta x_i \alpha_{i,steel} \tilde{h}_i^{steel,inner} \left( T_{i,1}^w - T_{steel} \right) \\ &+ \sum_{j=1}^{NJ} 2\pi r_j \Delta r_j \tilde{h}_{steelflow-bottom} \left( T_{NM,j}^b - T_{steel} \right) + \dot{Q}_{steel} \\ &+ \dot{M}_{steel,EAF} \left( C_{p,steel} T_{steel,EAF} - h_{steel} \right) \\ &+ \pi R^2 \tilde{h}_i^{slag,metal} \left( T_{slag} - T_{steel} \right) \\ &+ \sum_{k}^{N_{alloy}} \dot{m}_{alloy,k} \left\{ h_{alloy,k} \left( T_{feed} \right) - h_{steel} \right\} \end{aligned} \quad [40]$$

The first RHS sum represents the heat transfer along the vertical ladle wall, while the second summation term represents the heat transfer between steel and the bottom refractory. It is assumed that the bottom heat transfer is zero before steel has arrived in the ladle and is at first arrival time becoming non-zero (activated).

$\alpha_{i,steel}$ is the metal fraction contacting brick number i and varies with time. $\dot{M}_{steel,EAF}$ is the time dependent mass flow of steel coming from the EAF. $h_{steelflow-bottom}$ is the heat transfer coefficient for metal-bottom refractory heat exchange, and $\tilde{h}_i^{slag,metal}$ is the area averaged heat transfer coefficient between metal and slag. $\dot{Q}_{steel}$ is the heating power supplied directly to the steel [W/kg]. Again, these quantities are in general varying with time.

The heat source - and $\dot{Q}_{slag}$ are related to the total power $\dot{Q}_{tot}$ supplied by the heating electrodes. $\dot{Q}_{tot}$ is the power logged at the plant. The heat entering the slag and metal will be lower. We introduce and overall heating efficiency $\eta_{eff}$ [0,1] and a heat distribution coefficient $\eta_{slag}$, such that

$$\dot{Q}_{slag} = \frac{\eta_{slag} M_{slag} C_{p,slag}}{\eta_{slag} M_{slag} C_{p,slag} + M_{steel} C_{p,steel}} \eta_{eff} \dot{Q}_{tot} \quad [41]$$

$$\dot{Q}_{steel} = \frac{M_{steel}C_{p,steel}}{\eta_{slag}M_{slag}C_{p,slag} + M_{steel}C_{p,steel}} \eta_{eff}\dot{Q}_{tot} \qquad [42]$$

The coefficient $\eta_{slag} = 1.0$ tells that slag and metal increase temperature at the same rate. If $\eta_{slag} = 2.0$ the slags picks up temperature twice as fast as the steel. If $\eta_{slag} = 0.5$ the slag picks up temperature at half the rate of the steel. The introduction of the coefficient $\eta_{slag}$ allows a more controlled way to distribute heat addition between steel and slag.

*Solution for the energy equations*

Based on previous temperatures the radiation flows and fluxes are computed (equations [26]-[31]). For the radial wall elements the discrete equations [10], [13] and [19] can be written as

$$\mathbf{A_i} \cdot \mathbf{T}_i^{w,n+1} = \mathbf{b}_i^w, \qquad [43]$$

Here $\mathbf{b}_i^w$ will contain reference to previous slag and metal temperatures, radiation fluxes and external temperatures. The solution is obtained by inverting the NJxNJ (here 7x7) matrix $\mathbf{A_i}$:

$$\mathbf{T}_i^{w,n+1} = \mathbf{A_i^{-1}} \cdot \mathbf{b}_i^w \qquad [44]$$

We may notice that during the period when the ladle is in steady operation (no filling or tapping) the matrix $\mathbf{A_i}$ is fixed. In this case the new wall temperatures are obtained by only updating $\mathbf{b}_i^w$, which depends on values from previous time step, and then re-doing the matrix-vector operation in eq. [44]. This allows very fast solution of wall temperatures.

The bottom part of the wall is solved identically to what is explained above.

*Discrete equations for the slag and metal energy*

The coupled discrete equations for slag and metal enthalpy can be solved analytically, provided the inner refractory wall temperatures are known. First, we need to establish the relation between temperatures and enthalpies. This is elaborated in Appendix B Temperature-enthalpy relations. As seen from Appendix C Discrete equations for the slag-metal heat balance, explicit expressions for the slag and metal enthalpies are given by equation [92]. Temperatures are then computed by equations [75] and [77].

**Erosion model**

The erosion is primarily a result of dissolution and mass transfer from the refractory into the metal and slag. The erosion mechanisms considered are mass loss of refractory to the liquid by dissolution. In addition, we have mass losses due to thermal stresses. These may be addressed in a machine learning model, which may exploit the predicted difference between refractory temperature and incoming steel temperature.

### *Refractory loss in the steel wetted region*

During periods with considerable agitation on the metal and slag (bubble driven convection, natural convection, electromagnetic stirring) the carbon binder of the MgO-C refractory may be dissolved into the steel. The mass flux of carbon into the steel is locally given by:

$$\vec{J} = -\alpha_C D_C \rho_{steel} \nabla x_C \qquad [45]$$

Here $\alpha_C$ is the volume fraction of the refractor that is occupied by carbon, $D_C$ is the diffusivity of carbon in steel and $x_C$ is the mass fraction of carbon in the steel.

By introducing the concept of a mass transfer coefficient, we may write [45] as

$$\vec{J} = \alpha_C k_{C,BL} \rho_{steel} \left( x_C^{eq}(T_{wall}) - x_C^{bulk} \right) \vec{n} \qquad [46]$$

Here $k_{C,BL}$ is the mass transfer coefficient for the liquid side boundary layer and $x_C^{eq}(T_{wall})$ is the solubility of C into the steel with its actual composition, and where $T_{wall}$ is the temperature at the inner ladle wall. The temperature is controlled by the steel temperature and the temperature in the refractory brick. As the steel and the refractory has comparable thermal conductivities, the wall temperature will depend on both temperatures.

For forced convection we may use the mass transfer coefficient suggested by (Scalo et al., 2012) and (Shaw and Hanratty, 1977), stating that the mass transfer coefficient for Schmidt number Sc > 20 can be approximated by

$$k_{C,BL} = 0.09 \cdot u_\tau \cdot Sc^{-0.7} \qquad [47]$$

Values for the shear velocities are found in Typical values range from 0.0 to 0.1 m/s.

From equation [47] we learn that erosion of the steel wetted ladle wall will increase by gas stirring flow rate, increased temperature (increased C solubility and C diffusivity, decreased viscosity).

*Mass transfer resistance in the interface between MgO-C and steel*

At the inner surface of the MgO-C bricks the C binder will dissolve into the steel while MgO may be considered as inert. A sketch is provided in Figure 5. As the carbon binder is dissolved into the steel the average transport length $s_{pore}$ will stabilize around a typical MgO particle radius. If the MgO particles are small the convection inside the pores space can be neglected. In this case the transport in the pore space may be given by pure diffusion. In that case we may write:

$$\vec{J}_{porespace} = \alpha_C \frac{D_C}{s_{pore}} \rho_{steel} \left( x_C^{eq}(T_{wall}) - x_C^{IB} \right) \vec{n} \qquad [48]$$

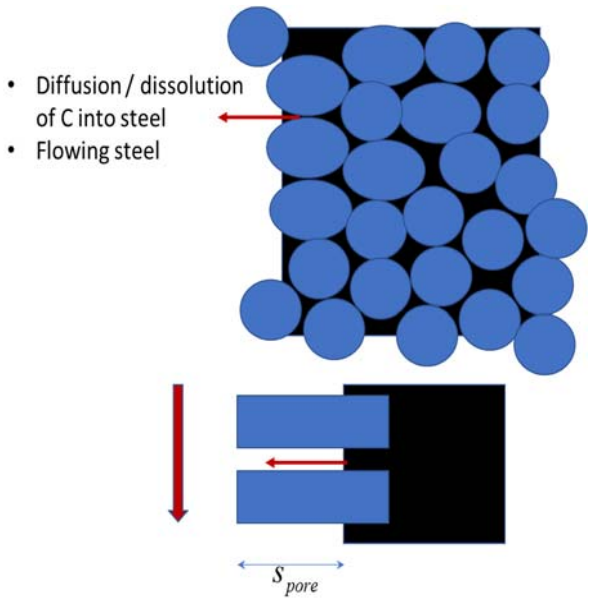

*Figure 5 Upper: MgO particle in a C matrix. The liquid flow of steel is on the LHS. Lower: Illustration of C that must diffuse through channels between MgO grains to reach the inner side of the flow boundary layer. The vertical arrow indicates the steel flow. Horizontal arrow indicates diffusion flux.*

Here $x_C^{IB}$ is the C mass fraction at the wall, defined at the outer surface made up if the MgO particles protruding out of the C matrix. In this case the mass flow through the inner and outer layers must match, giving:

$$\vec{J}_{eff} = \alpha_C k_{C,BL} \rho_{steel} \left( x_C^{IB} - x_C^{bulk} \right) \vec{n} = \alpha_C \frac{D_C}{s_{pore}} \rho_{steel} \left( x_C^{eq}(T_{wall}) - x_C^{IB} \right) \vec{n} \quad , \qquad [49]$$

And where the mass transfer coefficient is given by

$$k_{C,eff} = \frac{k_C D_C}{k_C s_{pore} + D_C} \qquad [50]$$

The effective mass transport of C from the MgO-C brick to the steel is then given by

$$\vec{J}_{eff} = \alpha_C k_{C,eff} \rho_{steel} \left( x_C^{eq}(T_{wall}) - x_C^{bulk} \right) \vec{n} \qquad [51]$$

### *Refractory loss in the slag wetted region*
The slag is collected at a relatively thin layer at the surface. Due to the bubble plume, case by the stirring gas, the slag will be pushed away from the plume and will gather close to the refractory wall. As the bubble plume is asymmetrically places, the slag thickness close to the refractory wall will vary along the ladle perimeter. We neglect these complexities and assume complete radial symmetry. The thickness $\delta_{slag}$ of the slag layer that contacts the refractory can be estimated by:

$$\delta_{slag} = \beta_{slag} M_{slag} / (\rho_{slag} \pi R (H_{steel})^2) \qquad [52]$$

The slag layer will move vertically, according to waves generated by the bubble plume, as illustrated in Figure 6. The slag layer has thickness $\delta_{slag}$ and wave amplitude $a_{wave}$.

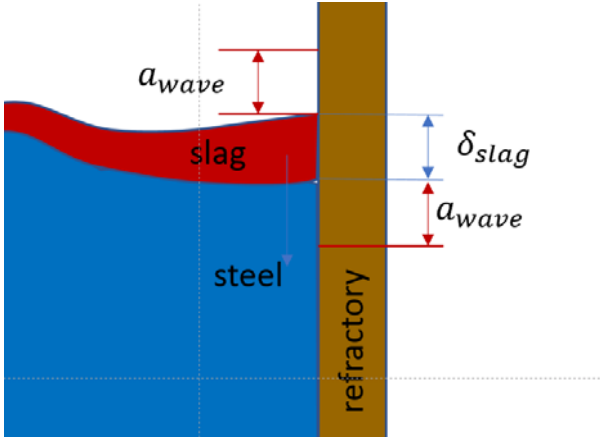

Figure 6 Illustration of the slag layer, close to the refractory, moving vertically with wave amplitude $a_{wave}$.

The mass transfer from wall to slag layer can be analyzed by assuming a developing boundary layer. According to Schlichting (Schlichting, 1979) the mass transfer along a developing boundary layer can be given by

$$Sh_x = \frac{kx}{D_{MgO}} = 0.339 \cdot Sc^{1/3} \sqrt{Re_x} \,, \qquad [53]$$

where k is the mass transfer coefficient and x is the distance along the developing boundary layer. $D_{MgO}$ is the diffusivity of MgO into the slag, and is related to the Schmidt number by

$$Sc = \frac{\nu_{slag}}{D_{MgO}} \qquad [54]$$

The explicit mass transfer coefficient is now:

$$k = \frac{\nu}{\sqrt{a_{wave} x}} 0.339 \cdot Sc^{-2/3} \sqrt{Re_{a_{wave}}} \qquad [55]$$

By averaging the mass transfer k in equation [55] over the thickness of the slag layer we can obtain

$$\bar{k} = 0.678 \cdot \frac{\nu_{slag}}{\delta_{slag}} Sc^{-2/3} \sqrt{\frac{u_{wave} a_{wave}}{\nu_{slag}}} \qquad [56]$$

The wave velocity $u_{wave}$ is now estimated by equation [108], and the swept distance (amplitude) $a_{wave}$ can be represented by $l_w$ in eq. [107]. It is possible to represent the distribution of mass transfer by a probability distribution. However, as a first approximation we may assume that the wave induced mass transfer applies to a region that extends over the thickness of the slag layer and a region that extends $a_{wave}$ both above and below the slag layer. In this case we may estimate the mass transport to the slag to be given over height $2a_{wave} + \delta_{slag}$, and where the average mass transfer coefficient for this layer is

$$k_{wave} = \bar{k} \frac{\delta_{slag}}{2a_{wave} + \delta_{slag}} = 0.678 \cdot \frac{v_{slag}}{2a_{wave} + \delta_{slag}} Sc^{-2/3} \sqrt{\frac{U_{wave} a_{wave}}{v_{slag}}} \quad [57]$$

In addition to the explicit wave contribution to mass transfer, the impact of the bubble driven flow (slag version of eq. [47]) must be added:

$$k_{eff} = k_{wave} + k_{MgO,BL} = 0.09 \cdot u_\tau \cdot Sc^{-0.7} + 0.678 \cdot \frac{v_{slag}}{2a_{wave} + \delta_{slag}} Sc^{-2/3} \sqrt{\frac{U_{wave} a_{wave}}{v_{slag}}} \quad [58]$$

*Overall refractory loss model*
We will track both the MgO and C components of the refractory. We may note that bottom erosion is not included in the model for now. The bottom is included due to it's impact the thermal balance (heat storage).

It is assumed that when C is dissolved from the bricks in the steel region a corresponding amount of MgO is released and will end up in the slag. It is assumed that the density of bricks is related to C and the corresponding MgO volume fractions ($\alpha_C, \alpha_{MgO}$) and phase densities ($\rho_C, \rho_{MgO}$) by

$$\rho_{brick} = \alpha_C \rho_C + \alpha_{MgO} \rho_{MgO} \quad , \quad [59]$$

where $\alpha_C + \alpha_{MgO} = 1$. The MgO loss mass, $M_{MgO}$, from a brick element during time dt, eroding a slice of thickness l, is

$$M_{MgO} = \dot{J}_{MgO} A \alpha_{MgO} \Delta t = l A \alpha_{MgO} \rho_{MgO} \rightarrow l = \frac{\dot{J}_{MgO} A \alpha_{MgO}}{A \alpha_{MgO} \rho_{MgO}} \Delta t \quad [60]$$

Here A is the total area, and $A\alpha_{MgO}$ is the partial area where MgO is contacting slag.

The corresponding loss of C is due to loss of MgO is then

$$M_C = \dot{J}_C A \alpha_C \Delta t = l A \alpha_C \rho_C \quad [61]$$

From the equations [60] and [61] we find that in the slag region the carbon flux out of the carbon part of the refractory wall is given by

$$j_C^{slag} = l \frac{A\alpha_C \rho_C}{A\alpha_C \Delta t} = \frac{j_{MgO}^{slag} A\alpha_{MgO}}{A\alpha_{MgO} \rho_{MgO}} \Delta t \frac{A\alpha_C \rho_C}{A\alpha_C \Delta t} = j_{MgO}^{slag} \frac{\rho_C}{\rho_{MgO}} \quad [62]$$

According to [62] the volume flows of the carbon and steel are equal. However, the surface areas are different due to the actual volume fractions. The mass flow of carbon, per surface area, to the liquid in the slag region is then

$$j_C^{slag} \alpha_C = j_{MgO}^{slag} \alpha_C \frac{\rho_C}{\rho_{MgO}} \quad [63]$$

Similarly, the loss of MgO in the steel region due to carbon dissolution is

$$j_{MgO}^{steel} \alpha_{MgO} = j_C^{steel} \alpha_{MgO} \frac{\rho_{MgO}}{\rho_C} \quad [64]$$

*Carbon balance*

The C (carbon) is lost from the refractory by two mechanisms, depending on if we are in the steel wetted or slag wetted zone.

$$\frac{d}{dt}\left(M_{steel} x_C^{steel}\right) = \sum_{i=1}^{NI} \alpha_{i,steel} \alpha_C A_i k_{C,eff} \rho_{steel} \left(x_C^{eq,steel}(T_{wall}) - x_C^{steel}\right)$$
$$+ \sum_{i=1}^{NI} \alpha_{i,slag} \alpha_{MgO} A_i \left(\alpha_C \frac{\rho_C}{\rho_{MgO}}\right) k_{MgO,eff}^n \rho_{slag} \left(x_{MgO}^{eq,slag}(T_{wall,i}, x_{composition}^{slag}) - x_{MgO}^{slag,n}\right) \quad [65]$$

Here the summation is over all the vertical refractory bricks. Here $\alpha_{i,steel}$ is the local steel fraction (varies with height in the ladle) and $\alpha_C$ is the carbon fraction in the refractory brick. $A_i = 2\pi R \Delta x_i$ is the local wall area.

*MgO balance*

The MgO is lost from the refractory, similar to the two mechanisms as above.

$$\frac{d}{dt}\left(M_{slag} x_{MgO}^{slag}\right) = \left((1-\alpha_C)\frac{\rho_{MgO}}{\rho_C}\right) \sum_{i=1}^{NI} \alpha_{i,steel} \alpha_C k_{C,eff} A_i \rho_{steel} \left(x_C^{eq,steel}(T_{wall}) - x_C^{steel,n}\right)$$
$$+ \sum_{i=1}^{NI} (1-\alpha_C) \alpha_{i,slag}^* k_{MgO,eff}^n A_i \rho_{slag} \left(x_{MgO}^{eq}(T_{wall,i}, x_{composition}^{slag}) - x_{MgO}^{slag,n}\right) \quad [66]$$

Here $\alpha_C$ is the volume fraction of carbon in the brick, while $(1-\alpha_C)$ is the MgO fraction. $\alpha^*_{i,slag}$ is the wave enhanced slag fraction, being in contact with the lining. As a first approach for $\alpha^*_{i,slag}$ we used $\alpha^*_{i,slag} = 0.25\alpha_{i-1,slag} + 0.5\alpha_{i,slag} + 0.25\alpha_{i+1,slag}$.

The left-hand terms are split and the effect of total mass change entered into the models. In the case of the slag we have:

$$\frac{d}{dt}\left(M_{slag} X_{MgO}\right) = M_{slag}\frac{d}{dt}\left(X_{MgO}\right) + X_{MgO}\frac{d}{dt}\left(M_{slag}\right) , \quad [67]$$

where the mass balance was given by eq. [5]. According to these equations we may write [66] as

$$M_{slag}\frac{d}{dt}\left(x^{slag}_{MgO}\right) = \left((1-\alpha_C)\frac{\rho_{MgO}}{\rho_C}\right)\sum_{i=1}^{NI}\alpha_{i,steel}\alpha_C k_{C,eff} A_i \rho_{steel}\left(x^{eq,steel}_C(T_{wall}) - x^{steel,n}_C\right)$$
$$+\sum_{i=1}^{NI}(1-\alpha_C)\alpha^*_{i,slag} k^n_{MgO,eff} A_i \rho_{slag}\left(x^{eq}_{MgO}(T_{wall,i}, x^{slag}_{composition}) - x^{slag,n}_{MgO}\right) \quad [68]$$
$$-x^{slag}_{MgO}\left(\dot{M}_{slag,EAF} + \sum_{k=1}^{N_{slag}}\dot{m}_{slag,k}\right)$$

where it is assumed that there is no MgO in the slag arriving from the EAF.

Similarly, the mass balance for carbon becomes

$$\frac{d}{dt}\left(M_{steel} x^{steel}_C\right) = \sum_{i=1}^{NI}\alpha_{i,steel}\alpha_C A_i k_{C,eff} \rho_{steel}\left(x^{eq,steel}_C(T_{wall}) - x^{steel}_C\right)$$
$$+\sum_{i=1}^{NI}\alpha_{i,slag}\alpha_{MgO} A_i \left(\alpha_C \frac{\rho_C}{\rho_{MgO}}\right) k^n_{MgO,eff} \rho_{slag}\left(x^{eq,slag}_{MgO}(T_{wall,i}, x^{slag}_{composition}) - x^{slag,n}_{MgO}\right) \quad [69]$$
$$-\left(\dot{M}_{steel,EAF} + \sum_{k=1}^{N_{alloy}}\dot{m}_{alloy,k}\right)x^{steel}_C + \dot{M}_{steel,EAF} x^{steel,EAF}_C$$

The solubility of MgO in the slag is given as (see acknowledgements) by

$$x^{eq,slag}_{MgO}(T_{wall,i}, x^{slag}_{composition}) \approx x^{eq,slag}_{MgO}(T_{wall,i}) =$$
$$= 0.1\cdot\min[(-4.34\cdot 10^5 + 514.3\cdot\widehat{T})/(1+100.74\cdot\widehat{T} - 0.041\cdot\widehat{T}^2); \quad [70]$$
$$50.0\cdot(9.025 - 4.427\cdot 10^{-3}\cdot\widehat{T} - 7.78\cdot 10^6/\widehat{T}^2) + (-598.7 + 0.2927\cdot\widehat{T} + 5.015\cdot 10^8/\widehat{T}^2)]$$

Here $\widehat{T}$ is the temperature in $°C$. As the slag composition is not known we use a temperature dependency which is approximate for 50 $wt\%CaO$, 10 $wt\%SiO2$, 2.5 $wt\%FeO$ and remaining is $wt\%Al2O3$.

### Developing sub models – a multi-scale approach

In the present approach we used CFD simulations (Johansen and Boysan, 1988) to obtain the shear stresses along the wall of the ladle. We did not include the effects of the slag. Using dynamic simulations with slag present more details could be added, and based on curve fitting or lookup tables, the data could have been plugged into the model. This would have improved the accuracy.

FACT SAGE calculations was performed for the solubility of MgO in the slag (see Acknowledgements). At the present time it was not possible to use this detailed information as we have no information of the slag composition when slag arrives into the ladle from the EAF (Electric Arc furnace). Based on this it was possible to close the model equations and realize the models.

### Software

The model was coded in python 3, using libraries numpy, pandas, math, pickle, scipy, and we used matplotlib and vtk for plotting and visualization. The basic version of the model is available on github.com, at address https://github.com/SINTEF/refractorywear. The model is licensed under the open source MIT license (https://opensource.org/licenses/MIT ).

### Tuning the model

In the table below we show the physical and thermodynamical data that was used.

*Table 1 Physical properties. Here T is temperature in °C and $X_c$ is mass fraction of carbon dissolved in the steel.*

|  | Density ($\rho$) [kg/m3] | Kinematic viscosity ($\nu$) [m2/s] | Thermal conductivity ($\lambda$) [W/mK] | Specific heat capacity ($C_p$) [J/kg K] | Diffusivity (D) [m2/s] |
|---|---|---|---|---|---|
| Slag | 3400 | 0.2e-5 | 10 | 500 | - |
| Steel<br>- $\rho$ :(Ceotto, 2013)<br>- $C_p$ :[1] | $8320 - 0.835(T-273.15) + (-83.2 + 8.35 \cdot 10^{-3}(T-273.15))X_c$ | 1.0e-6 | 15 | $C_p = 821.0 - 0.434 \cdot T + 0.000232 \cdot T^2$ | - |
| Wear brick | 3540 (3040) | - | 6 | 1500 | - |
| Durable brick | 2900 | - | 2.7 | 1500 | - |
| Outer brick | 2500 | - | 2.0 | 718 | - |
| Insulation | 300 | - | 0.1 | 900 | - |
| Steel casing | 7100 | - | 12 | 450 | - |

---

[1] https://www.setaram.com/application-notes/an372-heat-capacity-of-a-steel-betwwen-50c-and-1550c-liquid-state/

| Carbon/ carbon in steel | 2250 | - | - | - | $1.1 \cdot (1.0 + X_c/0.053) \cdot 10^{-8}$ |
|---|---|---|---|---|---|
| MgO in slag | - | - | - | - | $3.0 \cdot 10^{-9}$ |

*Table 2 Solubilities*

| Carbon solubility in steel: $x_C^{eq}(T_{wall})$ | 0.1 |
|---|---|
| MgO solubility in slag: $x_{MgO}^{eq,slag}(T_{wall})$ | See eq. [70] |

*Table 3 Tuning parameters*

| Parameter | Value |
|---|---|
| S_pore [m] | 4.8e-4 |
| Heater efficiency | 0.85 |
| Thermal conductivity: MgO bricks (wear bricks) | 5.0 |
| Thermal conductivity: Durable brick | 1.0 |
| Thermal conductivity: Outer brick | 1.0 |
| Thermal conductivity: Insulation | 1.0 |
| Thermal conductivity: Steel casing | 1.0 |
| Mass of casing | 2.0 |
| $C_{p,steel}$ | 1.0 |
| $\rho_{steel}$ | 1.0 |
| Mass transfer coefficient for slag, wave induced | 1.0 |

Unfortunately, detailed geometrical data and process data will not be given due to company confidentiality. In order to apply the model to single heats operational data from Sidenor was read. The static data included steel mass, time with steel in the ladle, temperature of the steel before leaving the EAF, and cyclic data for vacuum pressure, heating power, measured steel temperatures, gas flow rates, mass of additions and addition's composition, all versus time. The simulation was initiated at the time when the ladle was filled with liquid steel from the EAF and commenced after 2 hours. Once the casting process was finished, the ladle is considered as empty, but still losing heat.

As there is no data on the initial slag mass or composition, it was not possible to work with change in slag composition in the model. The initial slag mass was therefore assumed to be always 500 kg. Another consequence was that we had to assume constant solubilities of C in steel and MgO fraction in the slag. As a result the solubility of MgO in the slag only depends on temperature (se eq. [70]. Furthermore, all additions were assumed to contribute to the slag. This is acceptable if the alloy additions are of same order or smaller than the pure slag contribution. However, for special steels addition levels are significant and the model should be updated such that additions are transferred to the metal.

Different additions have different thermodynamic properties, such as melting temperature and melting enthalpy. As this information was by large unknown, we used the same melting temperature and heat of melting for all additions.

First, we tuned the steel temperature as a good thermal prediction was a prerequisite for the erosion model. At the beginning of each heat it was found that the initial temperature in many cases was a leftover from the previous heat, so we decided to use the temperature measured in the EAF, but which was decreased by 50 K due to heat loss during the tapping process. The heats where the initial temperature was non-existing or resulted in large temperature residuals, the initial temperature was corrected in an iterative manner until the residual for average relative temperature was below 20 K. The residual was computed from all measured values, except the first that was not reliable. In both Figure 7 and Figure 8 we see successful simulations, showing zero order residuals of 5 and 3 K, respectively. The first order residuals (RMSE) are similarly 7 and 5 K. In both cases the initial temperature was optimized, but for heat 206217 the "measured" initial steel temperature was quite close to the optimized initial temperature. To obtain these results, the thermal efficiency of the heater was reduced to 85 % and the thermal conductivity of the refractory bricks and insulation was significantly increased (see Table 1 and Table 3).

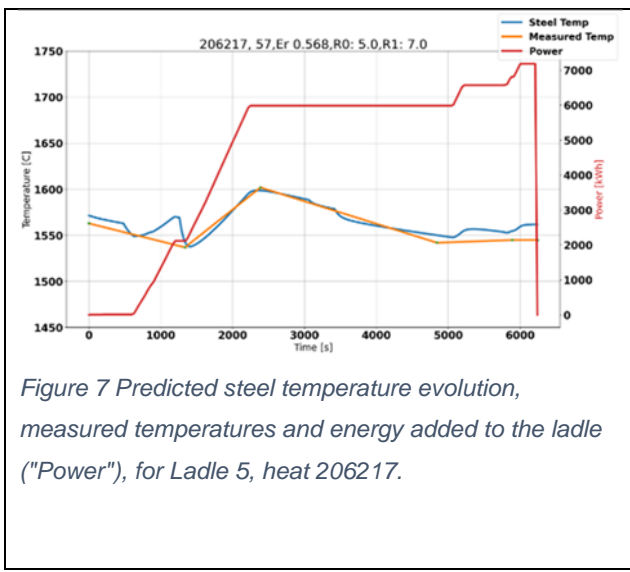

Figure 7 Predicted steel temperature evolution, measured temperatures and energy added to the ladle ("Power"), for Ladle 5, heat 206217.

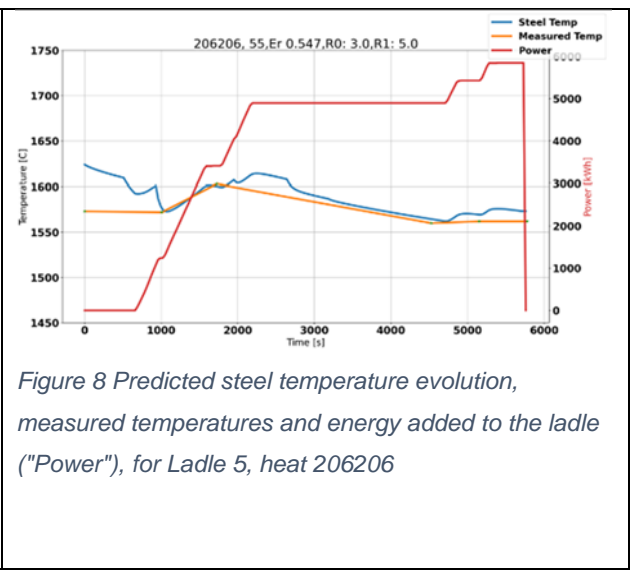

Figure 8 Predicted steel temperature evolution, measured temperatures and energy added to the ladle ("Power"), for Ladle 5, heat 206206

In the second step, the erosion model was tuned. We decided to work with constant solubilities of C in steel (soluble mass fraction was set to 0.1), while the MgO solubility in the slag is based on a fixed slag composition and only varies with temperature (see Table 2). As we decided to keep the solubility of C in the steel high constant, the only tuning that was possible is the pore diffusion length $s_{pore}$ (see eq. [48] and Figure 5).

To do this tuning we did following steps: i) Start with simulating the preheating of the ladle, ii) look up the heat ID, then read operational data for the heat and simulate temperature and erosion. iii) Based on the erosion data reduce the radial cell sizes for the three inner bricks (wear bricks), iv) Account for the thermal history of the ladle until next heat, v) Redo step ii) for the next use of the specific ladle (next heat in campaign, and where the campaign number is unique for the wear lining, from relining until demolition), and then accumulate the erosion of the bricks, vi) If the ladle was taken out for repair of some bricks, the repaired bricks are also repaired in the model. After repair the temperature is again initialized, vii) redoing step v) until the ladle is taken out for lining demolition. At this time the predicted erosion profiles are saved and compared to data from the demolition.

In the demolition data, the ladle is segmented in two halves, where "Left" is close to the porous plug while "Right" is away from the plug. In addition, the brick with most erosion in each half is registered. In this way, a maximum erosion is recorded and the average value for each brick row is not known. However, the 2D model can only be compared with the average of the two and should have some underprediction due to the above observation. For the selected tuning factor $s_{pore}$ we see that the prediction in Figure 9 is good, both qualitatively and quantitatively. The shape of the erosion in steel, below the slag line, is typical for all ladles and campaigns. We note that for bricks 36-40 the erosion level is quite high. This is above the liquid steel level and is a result of metal splashing, causing thermomechanical cracking, and disintegration due to the vacuum treatment (Jansson, 2008). In Figure 10 we see the prediction from a campaign where the erosion in the steel section (bricks 5-25) is underpredicted. This could be a result of the different steel qualities treated in this specific campaign or that for some reason the variation along the perimeter, at each brick layer, is larger than usual. As we have no data on the erosion from heat to heat, we cannot tell if this happened during specific heats in the campaign. Another interesting feature, seen in both Figure 9 and Figure 10, is the pronounced dip of erosion around brick 16 and 17. This may be a result of alloying materials addition when the ladle is approximately 1/3 full. Alloying elements and slag may stick to the colder wall long enough to protect the lining somewhat.

### Model performance against Sidenor operational data

The model was run with all available Sidenor data for 2019. The production campaigns that started in 2018 or ended in 2020 were omitted from the current data set as those campaigns were not complete. Altogether, we analyzed 5216 heats, involving 11 different ladles and 61 campaigns. Averaged erosion over bricks 5-25 is compared in Figure 11. An outlier (ladle 8, campaign 76), marked A, is seen and where the details were already shown in Figure 10. We compare the average erosion per heat in Figure 12, as distributed over the number of heats in each campaign. The model predicts a variation of ±12 %, while the data has a variation of ±18 %. The outlier A from Figure 11 is clearly seen.

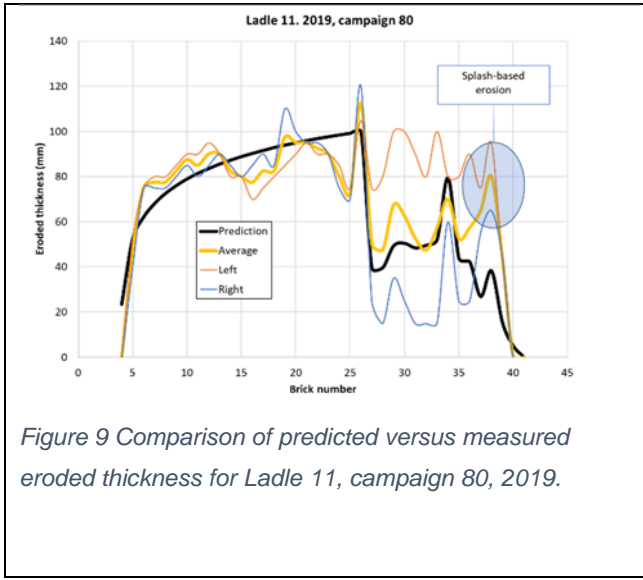

Figure 9 Comparison of predicted versus measured eroded thickness for Ladle 11, campaign 80, 2019.

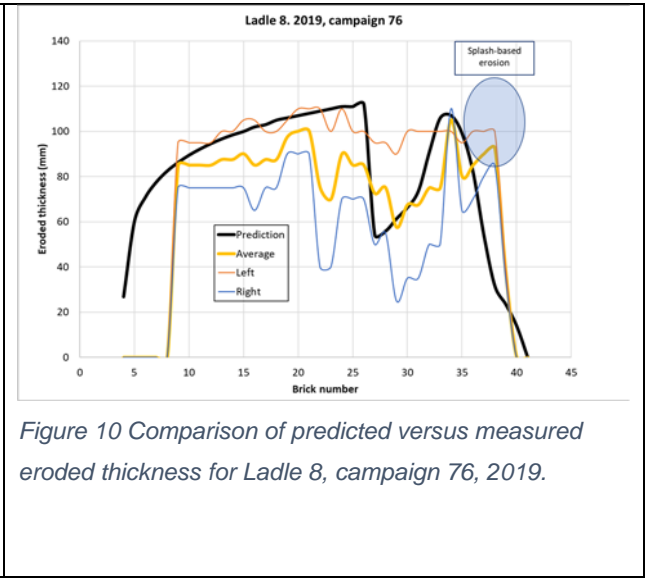

Figure 10 Comparison of predicted versus measured eroded thickness for Ladle 8, campaign 76, 2019.

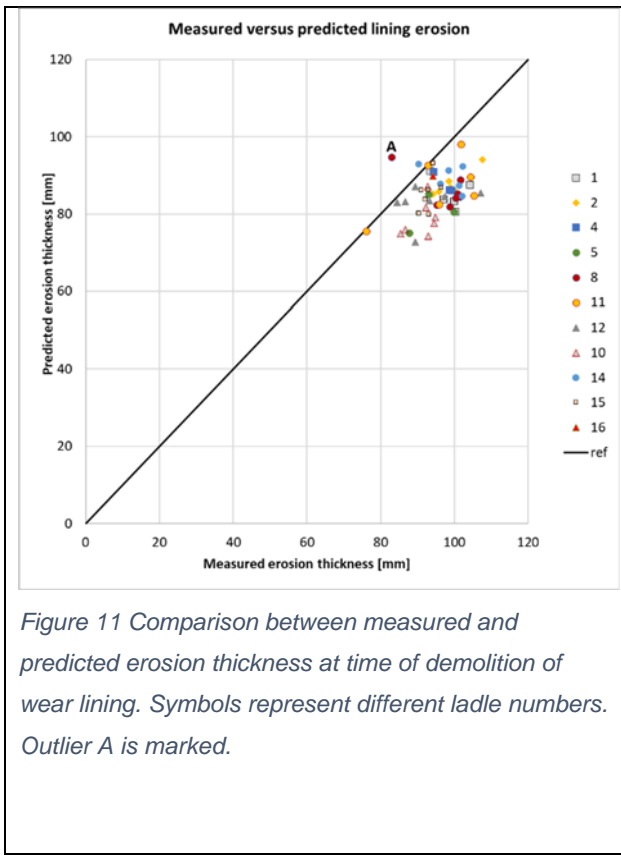

Figure 11 Comparison between measured and predicted erosion thickness at time of demolition of wear lining. Symbols represent different ladle numbers. Outlier A is marked.

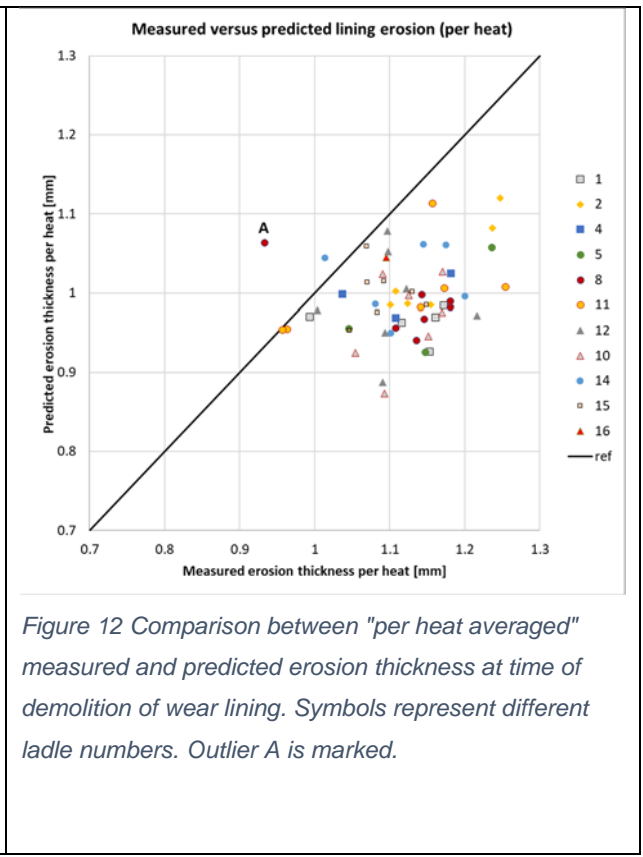

Figure 12 Comparison between "per heat averaged" measured and predicted erosion thickness at time of demolition of wear lining. Symbols represent different ladle numbers. Outlier A is marked.

It could be observed from Figure 9 and Figure 10 that a peak in erosion occurs close to the surface of the steel where the slag is located (around brick 35). The steel mass in the ladle varies from heat to heat, but in cases when the reported mass is low this may be due to operation challenges during casting. Therefore, the minimum steel mass is set to 110 tons. This introduces another uncertainty in the predictions. Now, it may seem that the erosion per heat does not change much from heat to heat, as indicated from Figure 12 We see in Figure 13 that the predictions show significant difference in amount eroded, and erosion pattern, from heat to heat. Around brick 25 (steel wetted region) the erosion for use number 17 is around twice as high than for use 69. This difference is mainly due to temperature, time with vacuum, gas flowrates and operational times. However, when averaged over a complete campaign, these variations are significantly reduced.

**Discussion**

The model predicts a smooth increase in erosion rate, from the bottom and towards the slag. This is in very good agreement with some of the measured erosion profiles. Figure 9 shows one example. This is a result of the bubble driven flow, enhanced by vacuum, the transport processes in the brick (represented by s_pore, or $s_{pore}$) and flow boundary layer, as well as the solubility of carbon in the steel. We used an artificially high value for the saturated carbon mass fraction ($X_C^{eq} = 0.1$). However, similar results as shown here may be obtained by another combination of $s_{pore}$ and $X_C^{eq}$.

We see above that the model performs quite well. At the same time there is room for improvements. The most obvious improvements are:

i) Modeling of the slag composition and adding the solubility of MgO in the slag as function of composition. However, this requires knowledge of the composition of the slag coming from the EAF.
ii) Separating additions into slag formers and alloy elements, and in addition update the enthalpy-temperature relations to represent the true composition of slag and metal.
iii) Empirical slag temperature is needed to calibrate and validate the slag temperature predictions.
iv) Including the solubility of carbon in the steel. Data for the steel composition is available but the carbon solubility for the different the compositions must be available.

Some features seen in the data, such as shown in *Figure* 14, cannot be reproduced by the model. The very high observed erosion rates, close to the bottom seems impossible to explain with the available information about the operation. A possible explanation could have been that gas purging was done with a very low steel level and containing slag. Such issues belong to the group of unnormal operations. Other possibilities are excessive mass loss during ladle cleaning, or that the lining brick quality was not consistent for a period.

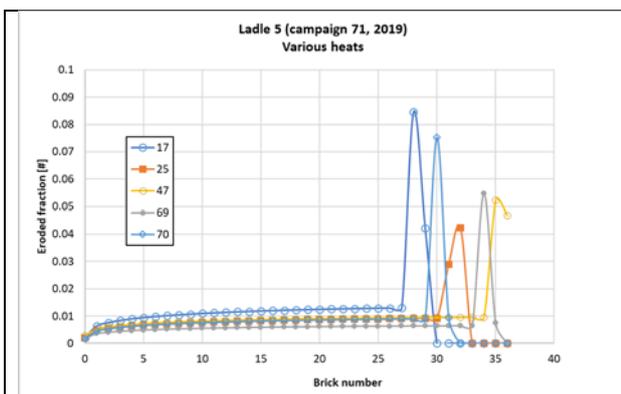

Figure 13 Predicted eroded fraction of wear lining versus brick number for 4 heats that are part of campaign 71. Graphs are labelled by ladle use number of the campaign. The symbols mark each brick, where a brick corresponds to a grid cell.

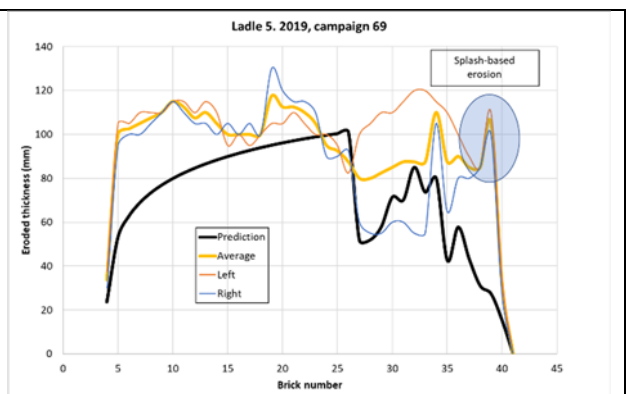

Figure 14 Erosion profile comparison for Ladle 5, campaign 69, 2019.

**Recommendations and conclusions**

The presented model predicts the evolution of the lining erosion fairly wall. Much better agreement between model and data is hard to obtain due to uncertainties in operational data, uncertainties in physical data and uncertainties in measurements. The model is primarily predicting lining erosion based on hydrodynamics and solution of lining elements in steel and slag. The contribution from thermomechanical cracking of the lining is not included in the model. However, the model predicts lining temperatures at the time of tapping metal into the ladle. This information can in the future be used to assess thermomechanical brick degradation. As this effect was not included the model was tuned to predict less erosion than what is observed. Similarly, the lining degradation above the melt, in particular pronounced during vacuum treatment, was not included in the model. However, a hole in the lining this far up on the ladle wall has far less consequences than holes deep below the steel surface.

Model predictions, as we have presented above, will be an important support for the ladle operator, when deciding if the ladle can be used one more time or not. The model shows how the variation in steel level between heats impacts erosion. If all the heats were run with same steel volume this would have a negative impact on lining lifetime. On the other side the refractory life may be extended by running scheduled amounts of steel in the heats. When the operator is unsure about the ladle conditions, and based on previous experience from running the model, the model prediction will help the operator to make a good decision.

The source code is made available to the public from https://github.com/SINTEF/refractorywear.

**Acknowledgements**

We thank Dr. Kai Tang, SINTEF Industry for his assistance with FACT SAGE calculations of MgO solubility in slag. The simplified MgO solubility versus temperature was based on this work.

This research was funded by the H2020 COGNITWIN project, which have received funding from the European Union's Horizon 2020 research and innovation program under grant agreement No. 870130.

**CRediT author statement**

STJ: Conceptualization, Methodology, Writing - Original draft preparation ; BTL: Software, Data curation, Validation, Visualization, Writing- Reviewing and Editing; TRD: Resources, Investigation, Writing- Reviewing and Editing.

**Nomenclature**

A      area [m$^2$]

p      pressure [Pa], or [bar]

V      volume [m$^3$]

M      mass [kg]

D      diffusivity [m$^2$/s]

k      mass transfer coefficient [m/s]

J      mass flux [kg/m$^2$s]

Nu      Nusselt number

Gr      Grashof number

Pr      Prandtl number

Ra      Rayleigh number

$\dot{m}_{alloy}$      mass flow of additions to the ladle during alloying and refining [kg/s]

$\dot{M}$      mass flow [kg/s]

Sc      Schmidt number

$s_{pore}$      pore diffusion length [m], also denoted s_pore, with ref to Figure 5

$u_\tau$      wall shear velocity [m/s]

| | |
|---|---|
| $U_{wave}$ | wave induced velocity [m/s], equation [108] |
| $\alpha$ | volume fraction |
| $\delta_{slag}$ | slag layer thickness [m] |
| $a_{wave}$ | wave amplitude [m] |
| $\beta$ | thermal expansion factor [1/K] |
| $\varepsilon$ | thermal emissivity [#] |
| $\rho$ | specific density [kg/m³] |
| $\mu$ | fluid viscosity [kg/ms] |
| $\nu$ | fluid kinematic viscosity [m2/s] |
| $\lambda$ | thermal conductivity [W/mK] |
| $\sigma$ | Stefan-Boltzmann coefficient (5.67e-8 W /m²K⁴) |
| $\tau_w$ | Wall shear stress [Pa] |
| $\Delta h$ | Specific heat of melting [kJ/kg] |
| $\Delta t$ | numerical time step [s] |
| $\Delta T$ | Change in temperature between two time steps: $\Delta T = T^{n+1} - T^n$ |
| $\Delta x, \Delta y$ | grid spacings in axial and radial directions [m] |
| $\Psi$ | Expression, given by equation [32] |
| $C_p$ | specific heat capacity [J/kg K] |
| h | specific enthalpy [kJ/kg] |
| $\tilde{h}$ | heat transfer coefficient [W/m²K] |
| H | height or distance [m] |
| $H_{steel}$ | total enthalpy of steel [kJ] |
| $H_{slag}$ | total enthalpy of slag [kJ] |
| r | radial position [m] |

| R | ladle inner radius [m] |
|---|---|
| T | temperature [K] |
| $x$ | mass fraction, subscripts representing carbon (c) or MgO |
| Q | gas volume flow rate [normal l/min] |
| $\dot{q}$ | heat flux [W/m$^2$] |
| $\dot{Q}$ | heat flow [W] |

Superscripts

| eq | thermodynamic equilibrium |
|---|---|
| EAF | Electric arc furnace |
| EXT | external |
| IB | inner boundary layer |

## Appendix A. Wall shear stress model

The steady state flow in the ladle was simulated (Johansen and Boysan, 1988) for the different (500, 600, 800 and 1200 Nl/min) argon gas flow rates, using a typical steel mass (130 tons) and ladle geometry from Sidenor. In addition, the simulations were performed for both atmospheric pressure and 0.003 bar above the melt. The simulated wall stresses were expressed by fitting functions and the effect of pressure was handled by a linear fit between the atmospheric and near vacuum pressures.

Have P as pressure in atmospheres ( P=1 bar = 1e5 Pa) , and we have volume flow of argon Q in units normal liters per min (nl/min), relative heigh x = X/H.

Wall stress data for low pressure (P=0.003 bar):

$$\tau_{w\_0.003bar}(x,Q) = ((-0.05201 + x \cdot 23.857)/(1 + x \cdot (-1.607) + x^2 \cdot 0.962))$$
$$\cdot (6.8377 + Q \cdot 0.02009)/(6.8377 + 1200 \cdot 0.02009)$$

Wall stress data for normal pressure (P=1.000 bar):

$$\tau_{w\_1.000bar}(x,Q) = ((-0.0736 + x \cdot 5.69)/(1 + x \cdot (-1.73) + x^2 \cdot 1.03))$$
$$\cdot (2.0563 + 0.005369 \cdot Q)/(2.0563 + 0.005369 \cdot 1200)$$

From the two fits above the wall stress distribution, including operating pressure, becomes

$$\tau_w(x,Q,P) = \tau_{w\_0.003bar}(x,Q)$$
$$+ \left( \tau_{w\_1.000bar}(x,Q) - \tau_{w\_0.003bar}(x,Q) \right) \cdot (P - 0.003)/(1 - 0.003)$$

[71]

Equation [71] is simply the new model. The result is shown in Figure 1515.

Wall shear velocity $u_\tau$ now becomes

$$u_\tau = \sqrt{|\tau_w(x,Q,P)|/\rho_{steel}}$$ [72]

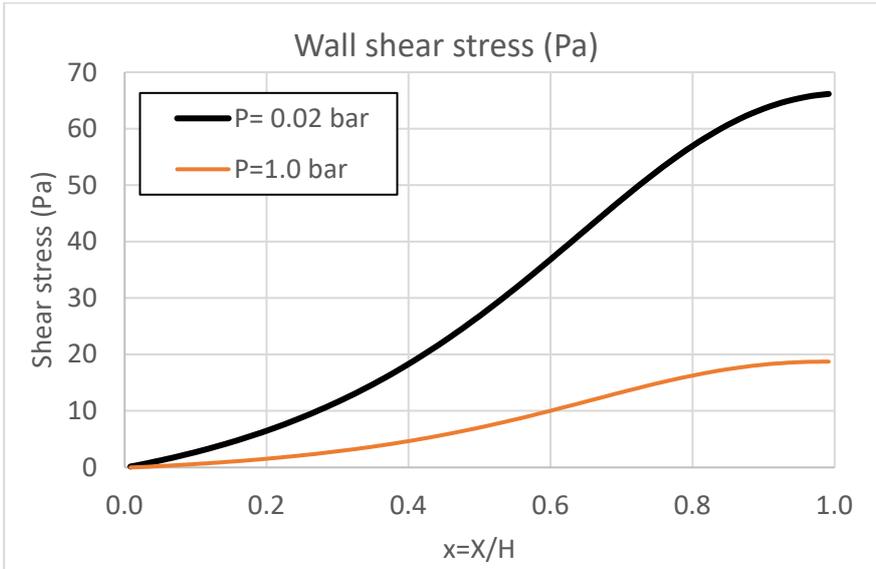

Figure 15 The distribution of shear stress (equation [71], *plotted for two absolute pressures above the melt.*

### Appendix B. Temperature-enthalpy relations

As we both use enthalpy and temperature, we establish some critical relations.

In case of the steel, we can find the temperature from the enthalpy by a Taylor expansion of the enthalpy function:

$$h_{steel}^{n+1} = h_{steel}(T^{n+1}) = h_{steel}(T^n) + \frac{h_{steel}(T^n + \Delta T) - h_{steel}(T^n)}{\Delta T}\left(T^{n+1} - T^n\right)$$ [73]

According to [73] the temperature is

$$T_{steel}^{n+1} = T_{steel}^n + \frac{h_{steel}^{n+1} - h_{steel}(T^n)}{h_{steel}(T^n + \Delta T) - h_{steel}(T^n)}\Delta T$$ [74]

As we wish to replace the temperatures of slag and metal with enthalpies, we do the following reorganizations:

$$T_{steel}^{n+1} = T_{steel}^n - \frac{h_{steel}(T^n)\Delta T}{h_{steel}(T^n+\Delta T)-h_{steel}(T^n)} + \frac{h_{steel}^{n+1}\Delta T}{h_{steel}(T^n+\Delta T)-h_{steel}(T^n)} \,,$$
$$= A_{steel} + B_{steel} h_{steel}^{n+1}$$
[75]

and where

$$A_{steel} = T_{steel}^n - \frac{h_{steel}(T^n)\Delta T}{h_{steel}(T^n+\Delta T)-h_{steel}(T^n)}$$
$$B_{steel} = \frac{\Delta T}{h_{steel}(T^n+\Delta T)-h_{steel}(T^n)}$$
[76]

Similarly, for the slag we get:

$$T_{slag}^{n+1} = T_{slag}^n - \frac{h_{slag}(T^n)\Delta T}{h_{slag}(T^n+\Delta T)-h_{slag}(T^n)} + \frac{h_{slag}^{n+1}\Delta T}{h_{slag}(T^n+\Delta T)-h_{slag}(T^n)} \,,$$
$$= A_{slag} + B_{slag} h_{slag}^{n+1}$$
[77]

where we have

$$B_{slag} = \frac{\Delta T}{h_{slag}(T^n+\Delta T)-h_{slag}(T^n)}$$
$$A_{slag} = T_{slag}^n - h_{slag}(T^n) B_{slag}$$
[78]

Based on relations [73] - [78] we can produce coupled enthalpy equations for the metal-slag-refractory system.

### Appendix C. Discrete equations for the slag-metal heat balance
*Fluid temperatures*

We can now write discrete equations for the slag and steel temperature. We use implicit treatment of the RHS enthalpies. The slag heat balance, given by equation [37] now reads

$$M_{slag} \frac{h_{slag}^{n+1}-h_{slag}^n}{\Delta t} = \sum_{i=1}^{NI} 2\pi R \Delta x_i \alpha_{i,slag} \tilde{h}_i^{slag,inner}\left(T_{i,1}^{w,n+1} - A_{slag} - B_{slag} h_{slag}^{n+1}\right) + \dot{Q}_{slag}$$
$$+ \dot{M}_{slag,EAF}\left(h_{slag,EAF} - h_{slag}^{n+1}\right)$$
$$+ \pi R^2 \tilde{h}_i^{slag,lid}(T_{lid} - A_{slag} - B_{slag} h_{slag}^{n+1})$$
$$+ \pi R^2 \tilde{h}_i^{slag,metal}(A_{steel} + B_{steel} h_{steel}^{n+1} - A_{slag} - B_{slag} h_{slag}^{n+1})$$
$$+ \sum_{k}^{N_{slag}} \dot{m}_{slag,k}\left\{h_{slag,k}(T_{feed}) - h_{slag}^{n+1}\right\}$$
[79]

while for the metal (steel) heat balance we have

$$M_{steel} \frac{h_{steel}^{n+1} - h_{steel}^n}{\Delta t} = \sum_{i=1}^{NI} 2\pi R \Delta x_i \alpha_{i,steel} \tilde{h}_i^{steel,inner} \left( T_{i,1}^w - A_{steel} - B_{steel} h_{steel}^{n+1} \right)$$

$$+ \sum_{j=1}^{NJ} 2\pi r_j \Delta r_j h_{steelflow-bottom} \left( T_{NM,j}^b - A_{steel} - B_{steel} h_{steel}^{n+1} \right) + \dot{Q}_{steel}$$

$$+ \dot{M}_{steel,EAF} \left( h_{steel,EAF} - h_{steel} \right) \qquad [80]$$

$$+ \pi R^2 \tilde{h}_i^{slag,metal} (A_{slag} + B_{slag} h_{slag}^{n+1} - A_{steel} - B_{steel} h_{steel}^{n+1})$$

$$+ \sum_{k}^{N_{alloy}} \dot{m}_{alloy,k} \left\{ h_{alloy,k}(T_{feed}) - h_{steel} \right\}$$

The discrete equation for the slag can be written in a simplified form:

$$h_{slag}^{n+1} \left( \frac{M_{slag}}{\Delta t} + \dot{M}_{slag,EAF} + \left[ \begin{array}{c} \pi R^2 \left( \tilde{h}_i^{slag,metal} + \tilde{h}_i^{slag,lid} \right) \\ + \sum_{k}^{N_{slag}} \dot{m}_{slag,k} + \sum_{i=1}^{NI} 2\pi R \Delta x_i \alpha_{i,slag} \tilde{h}_i^{slag,inner} \end{array} \right] B_{slag} \right)$$

$$= \frac{M_{slag}}{\Delta t} h_{slag}^n + \sum_{i=1}^{NI} 2\pi R \Delta x_i \alpha_{i,slag} \tilde{h}_i^{slag,inner} \left( T_{i,1}^{w,n+1} - A_{slag} \right) + \dot{Q}_{slag}$$

$$+ \dot{M}_{slag,EAF} h_{slag,EAF} + \pi R^2 \tilde{h}_i^{slag,lid} (T_{lid} - A_{slag}) \qquad [81]$$

$$+ \pi R^2 \tilde{h}_i^{slag,metal} (A_{steel} - A_{slag}) + \sum_{k}^{N_{slag}} \dot{m}_{slag,k} h_{slag,k}(T_{feed})$$

$$+ \pi R^2 \tilde{h}_i^{slag,metal} B_{steel} h_{steel}^{n+1}$$

We simplify eq. [81] as

$$h_{slag}^{n+1} E_{slag}^{n+1} = F_{slag}^n + G_{slag,wall}^n + H_{sm-steel}^n h_{steel}^{n+1} \qquad [82]$$

where

$$E_{slag}^{n+1} = \frac{M_{slag}}{\Delta t} + \dot{M}_{slag,EAF} + \left[ \begin{array}{c} \pi R^2 \left( \tilde{h}_i^{slag,metal} + \tilde{h}_i^{slag,lid} \right) \\ + \sum_{k}^{N_{slag}} \dot{m}_{slag,k} + \sum_{i=1}^{NI} 2\pi R \Delta x_i \alpha_{i,slag} \tilde{h}_i^{slag,inner} \end{array} \right] B_{slag} \quad, \qquad [83]$$

$$F_{slag}^n = \frac{M_{slag}}{\Delta t} h_{slag}^n + \dot{Q}_{slag}$$

$$+ \dot{M}_{slag,EAF} h_{slag,EAF} + \pi R^2 \tilde{h}_i^{slag,lid} (T_{lid} - A_{slag}) \quad, \qquad [84]$$

$$+ \pi R^2 \tilde{h}_i^{slag,metal} (A_{steel} - A_{slag}) + \sum_{k}^{N_{slag}} \dot{m}_{slag,k} h_{slag,k}(T_{feed})$$

$$G_{slag,wall}^{n} = \sum_{i=1}^{NI} 2\pi R \Delta x_i \alpha_{i,slag} \tilde{h}_i^{slag,inner} \left( T_{i,1}^{w,n+1} - A_{slag} \right) \quad , \text{ and} \quad [85]$$

$$H_{sm-steel}^{n} = \pi R^2 \tilde{h}_i^{slag,metal} B_{steel} \quad . \quad [86]$$

We have here treated the internal wall temperature in an explicit manner in order to be able to separate the set of equations and allow fast computations. The coefficients represented by the equations [83]-[86] have to be updated for every time step.

For the steel phase we may, based on eq. [80], write

$$h_{steel}^{n+1} E_{steel}^{n+1} = F_{steel}^{n} + G_{steel,walls}^{n} + H_{sm-slag}^{n} h_{slag}^{n+1} \quad , \quad [87]$$

where

$$E_{steel}^{n+1} = \frac{M_{steel}}{\Delta t} + B_{steel} \sum_{i=1}^{NI} 2\pi R \Delta x_i \alpha_{i,steel} \tilde{h}_i^{steel,inner} + B_{steel} \sum_{j=1}^{NJ} 2\pi r_j \Delta r_j h_{steelflow-bottom}$$
$$+ \dot{M}_{steel,EAF} + \pi R^2 \tilde{h}_i^{slag,metal} B_{steel} + \sum_{k}^{N_{alloy}} \dot{m}_{alloy,k} \quad , \quad [88]$$

and

$$F_{steel}^{n} = \frac{M_{steel}}{\Delta t} h_{steel}^{n} + \dot{Q}_{steel} + \dot{M}_{steel,EAF} \left( h_{steel,EAF} \right) + \sum_{k}^{N_{alloy}} \dot{m}_{alloy,k} \left\{ h_{alloy,k} \left( T_{feed} \right) \right\}$$
$$+ \pi R^2 \tilde{h}_i^{slag,metal} \left( A_{slag} - A_{steel} \right) \quad , \quad [89]$$

and

$$G_{steel,walls}^{n}$$
$$= \sum_{i=1}^{NI} 2\pi R \Delta x_i \alpha_{i,steel} \tilde{h}_i^{steel,inner} \left( T_{i,1}^{w,n} - A_{steel} \right) + \sum_{j=1}^{NJ} 2\pi r_j \Delta r_j h_{steelflow-bottom} \left( T_{NM,j}^{b.n} - A_{steel} \right) \quad , \quad [90]$$

and where

$$H_{sm-slag}^{n} = \pi R^2 \tilde{h}_i^{slag,metal} B_{slag} \quad [91]$$

The solution for the equations [82] and [87] is trivial, giving:

$$h_{slag}^{n+1} = \frac{E_{steel}^{n+1}\left(F_{slag}^{n}+G_{slag,wall}^{n}\right)+H_{sm-steel}^{n}\left(F_{steel}^{n}+G_{steel,walls}^{n}\right)}{E_{steel}^{n+1}E_{slag}^{n+1}-H_{sm-steel}^{n}H_{sm-slag}^{n}} \quad [92]$$

$$h_{steel}^{n+1} = \left[H_{sm-slag}^{n}h_{slag}^{n+1} + F_{steel}^{n} + G_{steel,walls}^{n}\right]/E_{steel}^{n+1}$$

The calculation of temperatures from the enthalpies are given by equations [75] - [78].

## Appendix D. Pure natural and effective convection heat transfer

According to Ede (Ede, 1967) the heat transfer due to pure natural convection can be estimated by:

$$Nu_x = \frac{3}{4}\left[\frac{2\Pr}{5(1+2\Pr^{1/2}+2\Pr)}\right]^{1/4}(Gr_x \Pr)^{1/4}, \quad [93]$$

where we have definitions:

$$\Pr = \frac{\mu C_p}{\lambda} \quad \text{(Prandtl number)} \quad [94]$$

$$Nu_x = \frac{\dot{q}_w}{T_w - T_e}\frac{x}{\lambda} \quad \text{(Nusselt number)} \quad [95]$$

$$Gr_x = \frac{g\beta x^3(T_w - T_e)}{\nu^2} \quad \text{(Grashof number)} \quad [96]$$

An average distance x may be estimated by the liquid height $H_{steel}$, giving $\overline{x^3} = \frac{1}{4}H_{steel}^3$,

and

$$\beta = -\frac{1}{\rho}\left(\frac{\partial \rho}{\partial T}\right) \quad \text{(thermal expansion)} \quad [97]$$

The Churchill-Thelen correlation (Churchill and Chu, 1975) gives:

$$Nu_x = \left[3.657^{1/2} + \left(\frac{Gr_x \Pr}{300\left[(1+(\frac{0.5}{\Pr})^{9/16}\right]^{16/9}}\right)^{1/6}\right]^2 \quad [98]$$

The latter correlation is fine for turbulence controlled natural convection. However, for $Ra = Gr_x \Pr$ smaller than $10^{10}$ the model given by equation [93] is very good. We therefore propose a mixture of the two models with a transition at $Ra = Gr_x \Pr = 2 \cdot 10^{10}$.

The assembled model then becomes:

$$\widetilde{Nu_x} = A + (B-A)/(1+\exp(-(Ra/Ra_0 - 1))), \qquad [99]$$

where $Ra_0 = 2 \cdot 10^{10}$ , [100]

and the Rayleigh number is

$$Ra = Gr_x \Pr \quad , \text{ and} \qquad [101]$$

$$A = \frac{3}{4}\left[\frac{2\Pr}{5(1+2\Pr^{1/2}+2\Pr)}\right]^{1/4} (Gr_x \Pr)^{1/4} \quad , \text{ and} \qquad [102]$$

$$B = \left(\frac{Gr_x \Pr}{300\left[(1+(\frac{0.5}{\Pr})^{9/16}\right]^{16/9}}\right)^{1/3} . \qquad [103]$$

See Figure 16 for comparison to with the base correlations.

The heat transfer coefficient due to natural convection is now:

$$h_{NC} = \widetilde{Nu_x} \frac{\lambda}{x} \qquad [104]$$

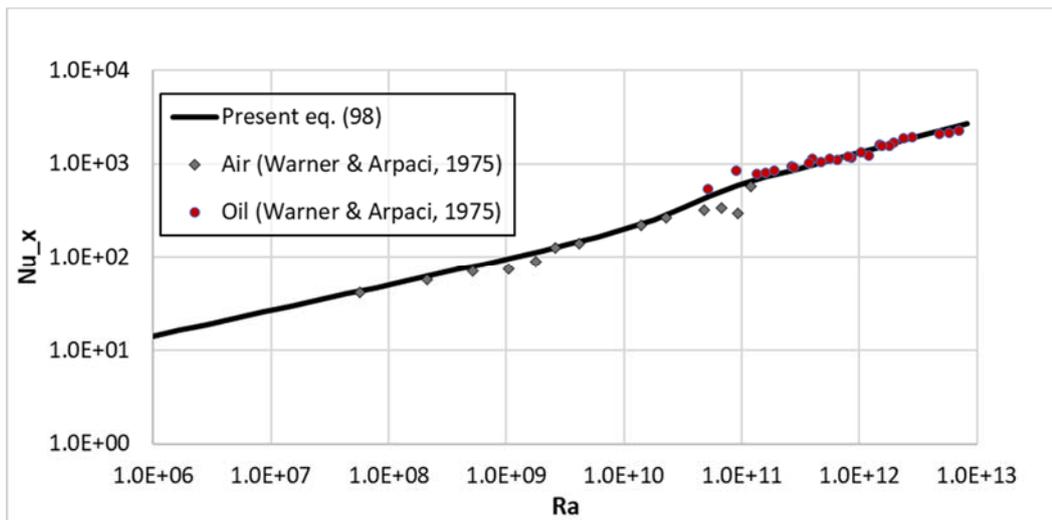

*Figure 16 The performance of the suggested natural convection heat transfer Nusselt number Nu<sub>x</sub>, eq. [99], plotted against Rayleigh number Ra. Data from Warner & Arpaci (Warner and Arpaci, 1968) is included*

*Effective heat transfer coefficient*

The effective heat transfer coefficient in the liquid steel and slag may now be estimated from

$$\tilde{h}_{liq} = \tilde{h}_{wave} + \left(\tilde{h}_{stirring}^{1/2} + \tilde{h}_{NC}^{1/2}\right)^2 \qquad [105]$$

## Appendix E Wave induced heat transfer

The bubble plume hitting the surface will produce waves at the interface (Cloete, Schalk. W. P:, 2008). An empirical correlation for the wave period $T_s$ was produced by (Hiratsuka et al., 2007)

$$\frac{(D/g)^{0.5}}{T_s} = \frac{0.459}{2\pi}\left(5.0 - \frac{H}{D}\right)^{0.5}\left(3.68\tanh\left(0.92\frac{H}{D}\left(5.0 - \frac{H}{D}\right)\right)\right)^{0.5}, \text{ where D is} \qquad [106]$$

ladle diameter and H is liquid height.

The length swept by the wave is $l_w$, where

$$l_w = \frac{U_{wave} T_s}{\pi}, \text{ and} \qquad [107]$$

The wave velocity is here based on the turbulent velocities near the free surface. From 2D CFD simulations we have curve-fitted the computed turbulent velocities to be given by:

$$U_{wave} = 0.0121 \cdot a^{\dot{Q}_g} \cdot \dot{Q}_g^{b}$$
$$a = 0.999925, \ b = 0.39233, \ [U_{wave}] = m/s, \ [\dot{Q}_g] = l/\min \qquad [108]$$

According to correlation [108] the wave velocities $U_{wave}$ will range between 0.13 and 0.18 m/s for gas flow rates ranging between 500 and 1200 l/min.

According to Blasius:

$$Nu_x = \frac{hx}{\lambda} = 0.332 \Pr^{0.33} \text{Re}_x^{0.5} \quad (0.5 < \Pr < 15) \qquad [109]$$

Over the distance $l_w / 2$ and average velocity $\bar{u} = \frac{2U_{wave}}{\pi}$, the averaged heat transfer coefficient $\tilde{h}_{wave}^*$ is:

$$\tilde{h}_{wave}^* = \frac{2\lambda}{l_w} Nu_{l_w} = \lambda \cdot 0.664 \Pr^{0.33}\left(\frac{\bar{u}l_w}{2\nu}\right)^{0.5} l_w^{-1} \quad (0.5 < \Pr < 15) \qquad [110]$$

The wave induced added heat transfer will apply to a region close to the surface. Due to the small thickness of the slag layer, it is proposed that the wave induced heat transfer applied only to the metal, given by the following relationship:

$$\tilde{h}_{wave} = \begin{cases} 0.0; & \dot{Q}_g = 0 \\ 0.0; & x > H_{metal} \\ \tilde{h}_{wave}^* \cdot e^{\frac{2\pi(H_{metal}-x)}{l_w}}; & x \leq H_{metal} \end{cases} \qquad [111]$$

**Appendix F Inner wall heat transfer coefficients due to forced convection by bubble stirring**

The heat transfer coefficients due to forced convection can be obtained from the definition of the dimensionless flow temperature $T^+$:

$$\dot{q}_w = \rho C_p u_\tau \frac{(T(y) - T_{wall})}{T^+(y^+, s^+, \Pr)} \qquad [112]$$

Here $\dot{q}_w$ is the wall heat flux, $u_\tau$ is the wall shear velocity, defined by $u_\tau = \sqrt{\frac{\tau_w}{\rho}}$, $y^+ = yu_\tau / \nu$ is the non-dimensional wall distance and $s^+ = su_\tau / \nu$ is the non-dimensional wall roughness.

The resulting heat transfer coefficient then becomes

$$\tilde{h}_{stirring} = \frac{\dot{q}_w}{(T(y) - T_{wall})} = \frac{\rho C_p u_\tau}{T^+(y^+, s^+, \Pr)} \qquad [113]$$

Using a wall function model (Ashrafian and Johansen, 2007) the $T^+$ function can be evaluated at a typical bulk fluid wall distance ($y^+ = 1000$, and for a given hydrodynamic roughness height s ( $s^+ = s \cdot u_\tau / \nu$ ), where $\nu$ is the kinematic viscosity of the steel.

$$T^+(y^+ = 1000, s^+, \Pr) = \frac{((5.95 + 13.6 \cdot \Pr^{0.596}) + (0.117 + 0.235 \cdot \Pr^{0.893}) \cdot s^+)}{(1 + (0.011 + 0.0939 \cdot \Pr^{0.676}) \cdot s^+ + (0.00005 + 0.0000683 \cdot \Pr^{0.62}) \cdot s^{+2})} \qquad [114]$$